\documentclass[12pt]{article}
\usepackage{amsmath}

\usepackage{amssymb}
\usepackage{bm}
\usepackage[dvips]{graphicx}

\newcommand{\llangle}{\langle\!\langle}
\newcommand{\rrangle}{\rangle\!\rangle}

\begin{document}

\baselineskip=17pt

\begin{titlepage}
\rightline{\tt LMU-ASC 30/15}
\rightline{\tt UT-Komaba/15-1}
\begin{center}
\vskip 2.5cm
{\Large \bf {$A_\infty$ structure from the Berkovits formulation}}\\
\vskip 0.4cm
{\Large \bf {of open superstring field theory}}
\vskip 1.0cm
{\large {Theodore Erler,${}^1$ Yuji Okawa${}^2$ and Tomoyuki Takezaki${}^2$}}
\vskip 1.0cm
${}^1${\it {Arnold Sommerfeld Center, Ludwig-Maximilians University}}\\
{\it {Theresienstrasse 37, 80333 Munich, Germany}}\\
tchovi@gmail.com
\vskip 0.5cm
${}^2${\it {Institute of Physics, The University of Tokyo}}\\
{\it {Komaba, Meguro-ku, Tokyo 153-8902, Japan}}\\
okawa@hep1.c.u-tokyo.ac.jp, takezaki@hep1.c.u-tokyo.ac.jp

\vskip 2.0cm

{\bf Abstract}
\end{center}

\noindent
By formulating open superstring field theory
based on the small Hilbert space of the superconformal ghost sector,
an action for the Neveu-Schwarz sector with an $A_\infty$ structure
has recently been constructed.
We transform this action to the Wess-Zumino-Witten-like form
and show that this theory is related to the Berkovits formulation
of open superstring field theory based on the large Hilbert space
by partial gauge fixing and field redefinition.

\end{titlepage}

\tableofcontents

\newpage

\section{Introduction}
\setcounter{equation}{0}

Can we consistently quantize string field theory?
This is one of the fundamental questions
when we work on string field theory.
While the gauge structure of string field theory is complicated,
open bosonic string field theory~\cite{Witten:1985cc}
and closed bosonic string field theory~\cite{Kaku:1988zv, Kaku:1988zw, Saadi:1989tb, Kugo:1989aa, Kugo:1989tk, Zwiebach:1992ie}
have been quantized based on the Batalin-Vilkovisky formalism~\cite{Batalin:1981jr, Batalin:1984jr}.
However, the quantization of the bosonic string
is formal because of the presence of tachyons.
How about the quantization of superstring field theory?

Among various formulations of superstring field theory,
the Berkovits formulation for the Neveu-Schwarz (NS) sector
of open superstring field theory~\cite{Berkovits:1995ab}
has been quite successful.
It is based on the large Hilbert space
of the superconformal ghost sector~\cite{Friedan:1985ey, Friedan:1985ge},
and various analytic solutions have been constructed~\cite{Erler:2007rh, Okawa:2007ri, Okawa:2007it, Fuchs:2007gw, Kiermaier:2007ki, Noumi:2011kn, Erler:2013wda}.
The construction of a classical master action in the Batalin-Vilkovisky formalism,
however, has turned out to be formidably complicated
for the Berkovits formulation~\cite{Kroyter:2012ni, Torii:2012nj, Torii:2011zz, Berkovits:2012np, Berkovits:201X}.
Why is it so complicated?

In bosonic string field theory,
the equation of motion and the gauge transformation can both be written
in terms of the same set of string products.
The string products satisfy the set of relations called
$A_\infty$~\cite{Stasheff:I, Stasheff:II, Getzler-Jones, Markl, Penkava:1994mu, Gaberdiel:1997ia}
for the open string
and $L_\infty$~\cite{Zwiebach:1992ie, Lada:1992wc, Schlessinger-Stasheff} for the closed string.
These structures play a crucial role in the Batalin-Vilkovisky quantization,
and they are closely related to the decomposition of the moduli space of Riemann surfaces.
The source of the difficulty for the Batalin-Vilkovisky quantization
in the Berkovits formulation can be seen in the free theory
by comparing the equation of motion and the gauge transformations.
The equation of motion takes the form
\begin{equation}
Q \eta \Phi = 0 \,,
\end{equation}
where $\Phi$ is the open superstring field in the large Hilbert space,
$Q$ is the BRST operator, and $\eta$ is the zero mode
of the superconformal ghost $\eta (z)$.
On the other hand, the gauge transformations are given by
\begin{equation}
\delta \Phi = Q \Lambda +\eta \Omega \,,
\end{equation}
where $\Lambda$ and $\Omega$ are the gauge parameters.
In the equation of motion, the product of $Q$ and $\eta$ appears,
while the sum of the term generated by $Q$
and the term generated by $\eta$ appears
in the gauge transformations.
This difference can be thought of as the source of the difficulty.
Working in the large Hilbert space obscures the relation
to the supermoduli space of super-Riemann surfaces,
and it might be one possible reason underlying the difficulty.
The approach to incorporating the Ramond sector into the Berkovits formulation
proposed in~\cite{Berkovits:2001im} is also complicated,
and it might also be related to our insufficient understanding
of the connection between the large Hilbert space
and the supermoduli space of super-Riemann surfaces.

If we formulate open superstring field theory
based on the small Hilbert space,
the equation of motion of the free theory is
\begin{equation}
Q \Psi = 0 \,,
\end{equation}
where $\Psi$ is the open superstring field in the small Hilbert space,
and the gauge transformation is given by
\begin{equation}
\delta \Psi = Q \Lambda \,,
\end{equation}
where $\Lambda$ is the gauge parameter.
The equation of motion and the gauge transformation
are both written in terms of the BRST operator $Q$,
and this seems to be promising for constructing
string products satisfying the $A_\infty$ relations
in the interacting theory.
In fact, the covering of the supermoduli space of super-Riemann surfaces
is more closely related to formulations based on the small Hilbert space.
It had long been thought, however, that a regular formulation
based on the small Hilbert space would be difficult
because of singularities coming from local picture-changing operators.

Recently, it was demonstrated in~\cite{Iimori:2013kha}
that a regular formulation of open superstring field theory
based on the small Hilbert space can be obtained from
the Berkovits formulation by partial gauge fixing.
In the process of the partial gauge fixing, an operator $\xi$
satisfying the relation
\begin{equation}
\{ \, \eta, \xi \, \} = 1
\end{equation}
is used, and we can realize such an operator
by a line integral of the supercoformal ghost $\xi (z)$.
For instance, we can choose the zero mode $\xi_0$ of $\xi (z)$
to be $\xi$.
While a line integral of $\xi (z)$ does not have simple transformation properties
under conformal transformations, the partial gauge fixing guarantees
that the resulting theory is gauge invariant.
The BRST transformation of the line integral of $\xi (z)$ yields
a line integral of the picture-changing operator,
and singularities associated with local insertions of the picture-changing operator
are avoided in this approach.
The equation of motion and the gauge transformation
can be systematically derived from those of the Berkovits formulation,
but it turned out that the resulting theory does not exhibit the $A_\infty$ structure.

Once we recognize that a line integral of $\xi (z)$ can be used
in constructing a gauge-invariant action of open superstring field theory,
we do not necessarily start from the Berkovits formulation.
In~\cite{Erler:2013xta}, an action with an $A_\infty$ structure
based on the small Hilbert space was constructed
using a line integral of $\xi (z)$ as a new ingredient.
Because of the $A_\infty$ structure of the theory,
the quantization based on the Batalin-Vilkovisky formalism is straightforward
just as in open bosonic string field theory.
The construction was further generalized to the NS sector of heterotic string field theory
and the NS-NS sector of type II superstring field theory in~\cite{Erler:2014eba}.

We now have two successful formulations
for the NS sector of open superstring field theory.
The Berkovits formulation is beautifully constructed
based on the large Hilbert space,
and the theory constructed in~\cite{Erler:2013xta}
is based on the small Hilbert space and exhibits the $A_\infty$ structure.
In this paper we show that the two theories are related
by partial gauge fixing and field redefinition.
In the rest of the introduction, we summarize the relation we found.

Let us start with the description of the Berkovits formulation.
We denote the open superstring field in the large Hilbert space by $\widetilde{\Phi}$.
The action takes the Wess-Zumino-Witten-like (WZW-like) form given by~\cite{Berkovits:1995ab}
\begin{equation}
S = {}-\frac{1}{2} \, \langle \, e^{-\widetilde{\Phi}} ( \eta e^{\widetilde{\Phi}} ),
e^{-\widetilde{\Phi}} ( Q e^{\widetilde{\Phi}} ) \, \rangle
-\frac{1}{2} \int_0^1 dt \, \langle \, e^{-\widetilde{\Phi} (t)} \partial_t e^{\widetilde{\Phi} (t)}, 
\{ \, e^{-\widetilde{\Phi} (t)} ( Q e^{\widetilde{\Phi} (t)} ), e^{-\widetilde{\Phi} (t)} ( \eta e^{\widetilde{\Phi} (t)} ) \, \} \, \rangle \,, 
\end{equation}
where $\widetilde{\Phi} (1) = \widetilde{\Phi}$, $\widetilde{\Phi} (0) = 0$,
$\langle \, A, B \, \rangle$ is the BPZ inner product of $A$ and $B$,
and all string products are defined by the star product~\cite{Witten:1985cc}.
It was shown in~\cite{Berkovits:2004xh} that this action can be written as
\begin{equation}
S = {}-\int_0^1 dt \, \langle \, \eta \, ( e^{-\widetilde{\Phi} (t)} \partial_t e^{\widetilde{\Phi} (t)} ), 
e^{-\widetilde{\Phi} (t)} ( Q e^{\widetilde{\Phi} (t)} ) \, \rangle \,,
\end{equation}
and it can be further transformed to
\begin{equation}
S = {}-\int_0^1 dt \, \langle \, A_t (t), 
Q A_\eta (t) \, \rangle
\label{action-A_eta-A_t}
\end{equation}
with
\begin{equation}
A_\eta (t) = ( \eta \, e^{\widetilde{\Phi} (t)} ) \, e^{-\widetilde{\Phi} (t)} \,, \qquad
A_t (t) = ( \partial_t e^{\widetilde{\Phi} (t)} ) \, e^{-\widetilde{\Phi} (t)} \,.
\label{INOT-parameterization}
\end{equation}
We will use this form of the action in this paper.
The dependence of $\widetilde{\Phi} (t)$ on $t$ is topological,
and the action is a functional of $\widetilde{\Phi}$, which is the value of $\widetilde{\Phi} (t)$ at $t=1$.
The action has nonlinear gauge invariances given by
\begin{equation}
A_\delta = Q \Lambda +D_\eta \, \Omega \,,
\end{equation}
where $\Lambda$ and $\Omega$ are gauge parameters and
\begin{equation}
A_\delta = ( \delta e^{\widetilde{\Phi}} ) \, e^{-\widetilde{\Phi}} \,, \qquad
D_\eta \, \Omega = \eta \Omega -( \eta e^{\widetilde{\Phi}} ) \, e^{-\widetilde{\Phi}} \, \Omega
-\Omega \, ( \eta e^{\widetilde{\Phi}} ) \, e^{-\widetilde{\Phi}} \,.
\end{equation}
All these properties of the action follow from the relations
\begin{equation}
\eta A_\eta (t) = A_\eta (t) A_\eta (t) \,, \qquad
\partial_t A_\eta (t) = \eta A_t (t) -A_\eta (t) A_t (t) +A_t (t) A_\eta (t)
\label{A_eta-A_t-relations}
\end{equation}
together with the fact that the cohomology of $\eta$ is trivial.

In~\cite{Iimori:2013kha}, the gauge invariance associated with $\Omega$ was used
to impose the following condition on $\widetilde{\Phi}$ for partial gauge fixing:
\begin{equation}
\xi \widetilde{\Phi} = 0 \,.
\end{equation}
The string field $\widetilde{\Phi}$ satisfying this condition can be written as
\begin{equation}
\widetilde{\Phi} = \xi \widetilde{\Psi}
\end{equation}
with $\widetilde{\Psi}$ in the small Hilbert space.
We replace $\widetilde{\Phi} (t)$ in the action
by $\xi \widetilde{\Psi} (t)$ satisfying
$\widetilde{\Psi} (1) = \widetilde{\Psi}$ and $\widetilde{\Psi} (0) = 0$ to obtain
\begin{equation}
S = {}-\int_0^1 dt \, \langle \, ( \partial_t e^{\xi \widetilde{\Psi} (t)} ) \, e^{-\xi \widetilde{\Psi} (t)}, 
Q ( ( \eta \, e^{\xi \widetilde{\Psi} (t)} ) \, e^{-\xi \widetilde{\Psi} (t)} ) \, \rangle \,.
\end{equation}
The dependence of $\widetilde{\Psi} (t)$ on $t$ is topological,
and we can regard this as an action of $\widetilde{\Psi}$ in the small Hilbert space.
The action is invariant under the residual gauge transformation
after the partial gauge fixing.

Now the string fields $A_\eta (t)$ and $A_t (t)$ satisfying~\eqref{A_eta-A_t-relations}
are parameterized by $\widetilde{\Psi} (t)$ in the small Hilbert space.
We show in this paper that the action 
with the $A_\infty$ structure constructed in~\cite{Erler:2013xta}
can be brought to the form~\eqref{action-A_eta-A_t},
where $A_\eta (t)$ and $A_t (t)$ satisfying~\eqref{A_eta-A_t-relations}
are parameterized differently in terms of $\Psi (t)$ in the small Hilbert space.
Since the relations~\eqref{A_eta-A_t-relations} are satisfied,
the dependence on $t$ is topological, and the action is a functional of $\Psi$,
which is the value of $\Psi (t)$ at $t=1$.
We show that the action in terms of $\Psi$ and the action in terms of $\widetilde{\Psi}$
are related by field redefinition and establish the equivalence of the theory
obtained by the partial gauge fixing of the Berkovits formulation
and the theory with the $A_\infty$ structure.

\begin{figure}
\begin{center}
\begin{picture}(370,140)
\thicklines
\put(110,100){\makebox(20,20){$\Phi$}}
\put(120,110){\circle{25}}
\put(140,115){\makebox(90,20){field redefinition}}
\put(185,110){\vector(1,0){45}}
\put(185,110){\vector(-1,0){45}}
\put(240,100){\makebox(20,20){$\widetilde{\Phi}$}}
\put(250,110){\circle{25}}
\put(0,60){\makebox(110,20){partial gauge fixing}}
\put(120,90){\vector(0,-1){40}}
\put(260,60){\makebox(110,20){partial gauge fixing}}
\put(250,90){\vector(0,-1){40}}
\put(110,20){\makebox(20,20){$\Psi$}}
\put(120,30){\circle{25}}
\put(240,20){\makebox(20,20){$\widetilde{\Psi}$}}
\put(250,30){\circle{25}}
\put(140,5){\makebox(90,20){field redefinition}}
\put(185,30){\vector(1,0){45}}
\put(185,30){\vector(-1,0){45}}
\end{picture}
\end{center}
\caption{The relation between the theory with the $A_\infty$ structure
in terms of $\Psi$ in the small Hilbert space
and the Berkovits formulation in terms of $\widetilde{\Phi}$ in the large Hilbert space.}
\label{relation-figure}
\end{figure}
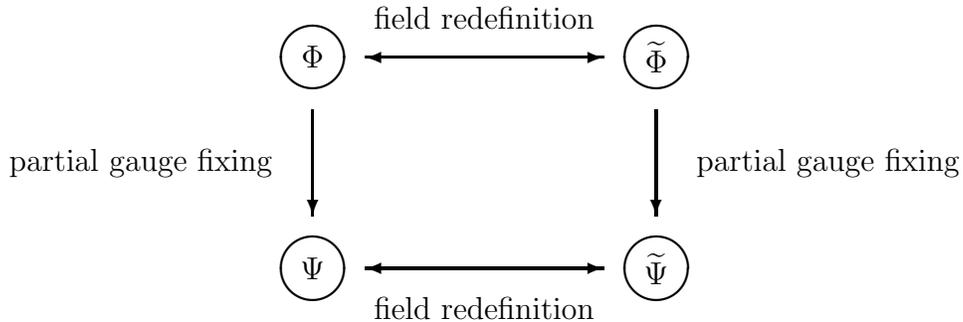

We further show that the action in terms of $\Psi (t)$
can be obtained from an action in terms of $\Phi (t)$ in the large Hilbert space
by the partial gauge fixing with the two string fields being related as $\Phi (t) = \xi \Psi (t)$.
Therefore, the theory constructed in~\cite{Erler:2013xta}
with the $A_\infty$ structure in terms of $\Psi$ in the small Hilbert space
can also be understood as being related
to the Berkovits formulation in terms of $\widetilde{\Phi}$
by field redefinition from $\widetilde{\Phi}$ to $\Phi$
and by the partial gauge fixing $\Phi = \xi \Psi$.
See figure~\ref{relation-figure}.

While transforming the action with the $A_\infty$ structure
constructed in~\cite{Erler:2013xta} to the WZW-like form is crucial
in showing its relation to the Berkovits formulation in this paper,
the equivalence of the theory with the $A_\infty$ structure
and the theory obtained from the Berkovits formulation by the partial gauge fixing
can also be shown from a different perspective.
In an accompanying paper~\cite{Erler:2015small},
the equivalence is shown by establishing that
the theory obtained from the Berkovits formulation by the partial gauge fixing
can be derived by a finite gauge transformation through the space of $A_\infty$ structures
just as in the case of the theory with the $A_\infty$ structure constructed in~\cite{Erler:2013xta}.

The rest of the paper is organized as follows.
In section~\ref{A_infinity-section}, we review
the construction of the $A_\infty$ structure in~\cite{Erler:2013xta}.
We begin in subsection~\ref{algebraic-ingredients}
by summarizing all the algebraic ingredients to be used in this paper.
We then briefly review the basics of the $A_\infty$ structure
in subsection~\ref{A_infinity-review}.
The construction of the $A_\infty$ structure in~\cite{Erler:2013xta}
is reviewed in subsection~\ref{A_infinity-theory},
and in this subsection we also present several new relations
which are essential in this paper.
In section~\ref{Berkovits-section}, we show that open superstring field theory
with the $A_\infty$ structure in the preceding section
is related to the Berkovits formulation
by partial gauge fixing and field redefinition.
We first transform the action with the $A_\infty$ structure
to the WZW-like form in subsection~\ref{WZW-like-form},
and we further write the action in a form with a topological dependence
on the parameter $t$ of the WZW-like action
in subsection~\ref{topological}.
We then establish its equivalence by field redefinition
to the theory we obtain from the Berkovits formulation by the partial gauge fixing
in subsection~\ref{field-redefinition},
and we embed the theory to the large Hilbert space in subsection~\ref{embedding}.
Section~\ref{conclusions-discussion} is devoted to conclusions and discussion.
In appendix~\ref{cohomomorphism}, we summarize a few identities
regarding cohomomorphisms.
In appendix~\ref{up-to-quartic}, we translate the $A_\infty$ conventions
into standard string field theory conventions, and we demonstrate
the equivalence of the action with the $A_\infty$ structure
and the action we obtain from the Berkovits formulation by the partial gauge fixing
up to quartic interactions explicitly.

\section{Open superstring field theory with the $A_\infty$ structure}
\label{A_infinity-section}
\setcounter{equation}{0}

\subsection{Algebraic ingredients}
\label{algebraic-ingredients}

In this subsection we present all the algebraic ingredients to be used in this paper.
We consider the NS sector of open superstring field theory.
An open superstring field is a state in the conformal field theory
which consists of the matter sector, the $bc$ ghost sector,
and the superconformal ghost sector.
We describe the superconformal ghost sector
using $\xi (z)$, $\eta (z)$, and $\phi (z)$~\cite{Friedan:1985ey, Friedan:1985ge}.
We denote the BPZ inner product of string fields $A$ and $B$
by $\langle \, A, B \, \rangle$.
It obeys
\begin{equation}
\langle \, B, A \, \rangle = (-1)^{AB} \langle \, A, B \, \rangle \,.
\end{equation}
Here and in what follows a state in the exponent of $-1$ represents its Grassmann parity:
it is $0$ mod $2$ for a Grassmann-even state and $1$ mod $2$ for a Grassmann-odd state.

Two important operators are the BRST operator which we denote by $Q$
and the zero mode of $\eta (z)$ which we denote by $\eta$.
They are Grassmann odd and obey
\begin{equation}
\begin{split}
& Q^2 = 0 \,, \qquad \eta^2 = 0 \,, \qquad \{ \, Q, \eta \, \} = 0 \,, \\
& \langle \, QA, B \, \rangle = {}-(-1)^A \, \langle \, A, QB \, \rangle \,, \qquad
\langle \, \eta A, B \, \rangle = {}-(-1)^A \, \langle \, A, \eta B \, \rangle \,.\,
\end{split}
\end{equation}

The cohomology of the operator $\eta$ is trivial
because there exists a Grassmann-odd operator $\xi$ which satisfies
\begin{equation}
\{ \, \eta, \xi \, \} = 1 \,.
\label{eta-xi}
\end{equation}
The existence of such an operator is crucial in this paper,
and we need to make a choice when we explicitly construct
various string products to be discussed.
However, the results of the paper are independent of the choice of $\xi$,
and we only use the relation~\eqref{eta-xi}.
When we make a choice of $\xi$, we use
a line integral of $\xi (z)$ to realize $\xi$,
and we further assume that
\begin{equation}
\xi^2 = 0 \,, \qquad 
\langle \, \xi A, B \, \rangle = (-1)^A \, \langle \, A, \xi B \, \rangle \,.\,
\end{equation}
For example, we can choose $\xi$ to be the zero mode $\xi_0$ of $\xi (z)$.

We say that a string field $A$ is in the small Hilbert space
when $A$ is annihilated by $\eta$:
\begin{equation}
\eta A = 0 \,.
\end{equation}
For a pair of string fields $A$ and $B$ in the small Hilbert space,
we define the BPZ inner product $\llangle \, A, B \, \rrangle$
in the small Hilbert space by
\begin{equation}
\llangle \, A, B \, \rrangle
= \langle \, \xi_0 A, B \, \rangle \,.
\end{equation}
It obeys
\begin{equation}
\llangle \, B, A \, \rrangle = (-1)^{AB} \llangle \, A, B \, \rrangle \,.
\end{equation}
We can use the operator $\xi$ to relate the two BPZ inner products as
\begin{equation}
\llangle \, A, B \, \rrangle
= \langle \, \xi A, B \, \rangle \,.
\end{equation}

Products of string fields are defined by the star product~\cite{Witten:1985cc}.
The star product is associative:
\begin{equation}
( A_1 A_2 ) \, A_3 = A_1 \, ( A_2 A_3 ) \,.
\label{associativity}
\end{equation}
The operators $Q$ and $\eta$ act as derivations with respect to the star product:
\begin{equation}
Q ( AB) = ( QA ) B +(-1)^A A (QB) \,, \qquad
\eta ( AB) = ( \eta A ) B +(-1)^A A (\eta B) \,.
\label{star-product-derivation}
\end{equation}
These are all the ingredients we will use in this paper.

\subsection{$A_\infty$ structure}
\label{A_infinity-review}

We briefly review the basics of the $A_\infty$ structure in this subsection.
See~\cite{Erler:2015small} for more comprehensive explanations.

Consider a two-string product $V_2 ( A_1, A_2 )$
and a three-string product $V_3 ( A_1, A_2, A_3)$
with
\begin{equation}
\epsilon ( V_2 ( A_1, A_2 ) ) = \epsilon (A_1) +\epsilon (A_2) \,, \qquad
\epsilon ( V_3 ( A_1, A_2, A_3 ) ) = \epsilon (A_1) +\epsilon (A_2) +\epsilon (A_3) +1 \,,
\end{equation}
where $\epsilon (A)$ denotes the Grassmann parity of $A$.
We say that $V_2 ( A_1, A_2 )$ is a Grassmann-even product
and $V_3 ( A_1, A_2, A_3)$ is a Grassmann-odd product.
They are assumed to have the following cyclic properties:
\begin{equation}
\begin{split}
\llangle \, A_1, V_2 ( A_2, A_3 ) \, \rrangle
& = \llangle \, V_2 ( A_1, A_2 ), A_3 \, \rrangle \,, \\
\llangle \, A_1, V_3 ( A_2, A_3, A_4 ) \, \rrangle
& = {}-(-1)^{A_1} \llangle \, V_3 ( A_1, A_2, A_3 ), A_4 \, \rrangle \,,
\end{split}
\label{V_2-V_3-cyclicity}
\end{equation}
where $A_1$, $A_2$, $A_3$, and $A_4$ are string fields in the small Hilbert space.
Consider an action given by
\begin{equation}
S = {}-\frac{1}{2} \, \llangle \, \Psi, Q \Psi \, \rrangle
-\frac{g}{3} \, \llangle \, \Psi, V_2 ( \Psi, \Psi ) \, \rrangle
-\frac{g^2}{4} \, \llangle \, \Psi, V_3 ( \Psi, \Psi, \Psi ) \, \rrangle
+O(g^3) \,,
\end{equation}
where $\Psi$ is a Grassmann-odd string field in the small Hilbert space
and $g$ is the coupling constant.
Because of the cyclic properties in~\eqref{V_2-V_3-cyclicity},
the variation of the action is given by
\begin{equation}
\delta S = {}-\llangle \, \delta \Psi, Q \Psi \, \rrangle
-g \, \llangle \, \delta \Psi, V_2 ( \Psi, \Psi ) \, \rrangle
-g^2 \, \llangle \, \delta \Psi, V_3 ( \Psi, \Psi, \Psi ) \, \rrangle
+O(g^3) \,,
\end{equation}
and the action is invariant up to $O(g^3)$
under the gauge transformation given by
\begin{equation}
\begin{split}
\delta_\Lambda \Psi & = Q \Lambda
+g \, \bigl( \, V_2 ( \Psi, \Lambda ) -V_2 ( \Lambda, \Psi ) \, \bigr) \\
& \quad~
+g^2 \, \bigl( \, V_3 ( \Psi, \Psi, \Lambda ) -V_3 ( \Psi, \Lambda, \Psi )
+V_3 ( \Lambda, \Psi, \Psi ) \, \bigr) +O(g^3)
\end{split}
\end{equation}
if $V_2 ( A_1, A_2 )$ and $V_3 ( A_1, A_2, A_3 )$ satisfy
\begin{equation}
\begin{split}
& Q V_2 ( A_1, A_2 ) -V_2 ( Q A_1, A_2 ) -(-1)^{A_1} V_2 ( A_1, Q A_2 ) = 0 \,, \\
& Q V_3 ( A_1, A_2, A_3 )
-V_2 ( V_2 ( A_1, A_2 ), A_3 ) +V_2 ( A_1, V_2 ( A_2, A_3 ) ) \\
& +V_3 ( Q A_1, A_2, A_3 ) +(-1)^{A_1} V_3 ( A_1, Q A_2, A_3 ) +(-1)^{A_1+A_2} V_3 ( A_1, A_2, Q A_3 )
= 0
\label{V_2-V_3-A_infinity}
\end{split}
\end{equation}
for states $A_1$, $A_2$, and $A_3$ in the small Hilbert space.

These relations are part of the $A_\infty$ relations.
To describe the $A_\infty$ structure, it is convenient to introduce {\it degree}
for a string field $A$ denoted by ${\rm deg} (A)$. It is defined by
\begin{equation}
{\rm deg} (A) = \epsilon (A) +1 \mod 2 \,,
\end{equation}
where $\epsilon (A)$ is the Grassmann parity of $A$.
We define $\omega ( A_1, A_2 )$, $M_2 ( A_1, A_2 )$, $M_3 ( A_1, A_2, A_3 )$ by
\begin{equation}
\begin{split}
\omega ( A_1, A_2 ) & = (-1)^{{\rm deg} (A_1)} \llangle \, A_1, A_2 \, \rrangle \,, \\
M_2 ( A_1, A_2 ) & = (-1)^{{\rm deg} (A_1)} \, V_2 ( A_1, A_2 ) \,, \\
M_3 ( A_1, A_2, A_3 ) & = (-1)^{{\rm deg} (A_2)} \, V_3 ( A_1, A_2, A_3 ) \,.
\end{split}
\end{equation}
Note that the sign factor for $M_3 ( A_1, A_2, A_3 )$ is different.
The inner product $\omega ( A_1, A_2 )$ is graded antisymmetric:
\begin{equation}
\omega ( A_1, A_2 )
= {}-(-1)^{{\rm deg} (A_1) \, {\rm deg} (A_2)} \, \omega ( A_2, A_1 ) \,.
\end{equation}
We have
\begin{equation}
\begin{split}
{\rm deg} ( Q A_1 )
& = {\rm deg} (A_1) +1 \,, \\
{\rm deg} ( M_2 ( A_1, A_2 ) )
& = {\rm deg} (A_1)+{\rm deg} (A_2) +1 \,, \\
{\rm deg} ( M_3 ( A_1, A_2, A_3 ) )
& = {\rm deg} (A_1)+{\rm deg} (A_2)+{\rm deg} (A_3) +1 \,,
\end{split}
\end{equation}
and we say that $Q$, $M_2$, and $M_3$ are degree odd.
The BPZ property of $Q$
and the cyclic properties~\eqref{V_2-V_3-cyclicity} are translated into
\begin{equation}
\begin{split}
\omega ( A_1, Q A_2 )
& = {}-(-1)^{{\rm deg} (A_1)} \, \omega ( Q A_1, A_2 ) \,, \\
\omega ( A_1, M_2 ( A_2, A_3 ) )
& = {}-(-1)^{{\rm deg} (A_1)} \, \omega ( M_2 ( A_1, A_2 ), A_3 ) \,, \\
\omega ( A_1, M_3 ( A_2, A_3, A_4 ) )
& = {}-(-1)^{{\rm deg} (A_1)} \, \omega ( M_3 ( A_1, A_2, A_3 ), A_4 ) \,,
\end{split}
\end{equation}
and the $A_\infty$ relations~\eqref{V_2-V_3-A_infinity} are written as
\begin{equation}
\begin{split}
& Q M_2 ( A_1, A_2 ) +M_2 ( Q A_1, A_2 ) +(-1)^{{\rm deg} (A_1)} M_2 ( A_1, Q A_2 ) = 0 \,, \\
& Q M_3 ( A_1, A_2, A_3 )
+M_2 ( M_2 ( A_1, A_2 ), A_3 ) +(-1)^{{\rm deg} (A_1)} M_2 ( A_1, M_2 ( A_2, A_3 ) ) \\
& +M_3 ( Q A_1, A_2, A_3 ) +(-1)^{{\rm deg} (A_1)} M_3 ( A_1, Q A_2, A_3 )
+(-1)^{{\rm deg} (A_1) +{\rm deg} (A_2)} M_3 ( A_1, A_2, Q A_3 ) = 0 \,.
\end{split}
\end{equation}

We construct multi-string products based on the star product.
Associated with the star product, we define $m_2 ( A_1, A_2 )$ by
\begin{equation}
m_2 ( A_1, A_2 ) = (-1)^{{\rm deg} (A_1)} A_1 A_2 \,.
\end{equation}
The two-string product $m_2$ is degree odd,
\begin{equation}
{\rm deg} ( m_2 ( A_1, A_2 ) )
= {\rm deg} (A_1)+{\rm deg} (A_2) +1 \,,
\end{equation}
and it has the following cyclic property:
\begin{equation}
\omega ( A_1, m_2 ( A_2, A_3 ) )
= {}-(-1)^{{\rm deg} (A_1)} \, \omega ( m_2 ( A_1, A_2 ), A_3 ) \,.
\end{equation}
The associativity of the star product~\eqref{associativity}
and the derivation property of $Q$ and $\eta$
with respect to the star product~\eqref{star-product-derivation}
can be stated in terms of $m_2$ as
\begin{equation}
\begin{split}
m_2 ( m_2 ( A_1, A_2 ), A_3 ) +(-1)^{{\rm deg} (A_1)} \, m_2 ( A_1, m_2 ( A_2, A_3 ) ) & = 0 \,, \\
Q \, m_2 ( A_1, A_2 ) +m_2 ( Q A_1, A_2 ) +(-1)^{{\rm deg} (A_1)} \, m_2 ( A_1, Q A_2 ) & = 0 \,, \\
\eta \, m_2 ( A_1, A_2 ) +m_2 ( \eta A_1, A_2 ) +(-1)^{{\rm deg} (A_1)} \, m_2 ( A_1, \eta A_2 ) & = 0 \,.
\end{split}
\label{m_2-properties}
\end{equation}

For the description of the $A_\infty$ structure,
it is further convenient to consider linear operators
acting on the vector space $T \mathcal{H}$ defined by
\begin{equation}
T \mathcal{H}
= \mathbb{C} \, \oplus \mathcal{H}
\oplus \mathcal{H}^{\otimes 2} \oplus \mathcal{H}^{\otimes 3} \oplus  \ldots \,,
\end{equation}
where $\mathbb{C}$ is the space of complex numbers,
$\mathcal{H}$ is the state space of the boundary conformal field theory,
and $\mathcal{H}^{\otimes n}$ is the space obtained by tensoring $n$ copies of $\mathcal{H}$.
For example, we define ${\bf Q}$ acting on $T \mathcal{H}$
associated with the BRST operator $Q$, which is degree odd, as follows:
\begin{equation}
\begin{split}
{\bf Q} \, 1 & = 0 \,, \\
{\bf Q} \, A_1 & = Q A_1 \,, \\
{\bf Q} \, ( A_1 \otimes A_2 )
& = Q A_1 \otimes A_2 +(-1)^{{\rm deg} (A_1)} A_1 \otimes Q A_2 \,, \\
{\bf Q} \, ( A_1 \otimes A_2 \otimes A_3 )
& = Q A_1 \otimes A_2 \otimes A_3
+(-1)^{{\rm deg} (A_1)} A_1 \otimes Q A_2 \otimes A_3 \\
& \quad~ +(-1)^{{\rm deg} (A_1) +{\rm deg} (A_2)} A_1 \otimes A_2 \otimes Q A_3 \,, \\
& \quad \vdots
\end{split}
\end{equation}
With this definition, we find
\begin{equation}
{\bf Q}^2 = 0 \,.
\label{Q^2=0}
\end{equation}
We similarly define ${\bm \eta}$ for $\eta$, which is also degree odd, and find
\begin{equation}
{\bm \eta}^2 = 0 \,, \qquad {\bf Q} \, {\bm \eta} +{\bm \eta} \, {\bf Q} = 0 \,.
\label{Q-eta}
\end{equation}
We denote the commutator of ${\bf C}_1$ and ${\bf C}_2$ graded with respect to degree
by $[ \, {\bf C}_1, {\bf C}_2 \, ]$.
Then the relations in~\eqref{Q^2=0} and~\eqref{Q-eta} can be written as
\begin{equation}
[ \, {\bf Q}, {\bf Q} \, ] = 0 \,, \qquad
[ \, {\bm \eta}, {\bm \eta} \, ] = 0 \,, \qquad
[ \, {\bf Q}, {\bm \eta} \, ] = 0 \,.
\end{equation}
For the two-string product $M_2 ( A_1, A_2 )$
and the three-string product $M_3 ( A_1, A_2, A_3 )$,
which are both degree odd, we define ${\bf M}_2$ and ${\bf M}_3$, respectively, by
\begin{equation}
\begin{split}
{\bf M}_2 \, 1 & = 0 \,, \\
{\bf M}_2 \, A_1 & = 0 \,, \\
{\bf M}_2 \, ( A_1 \otimes A_2 )
& = M_2 ( A_1, A_2 ) \,, \\
{\bf M}_2 \, ( A_1 \otimes A_2 \otimes A_3 )
& = M_2 ( A_1, A_2 ) \otimes A_3
+(-1)^{{\rm deg} (A_1)} A_1 \otimes M_2 ( A_2, A_3 ) \,, \\
{\bf M}_2 \, ( A_1 \otimes A_2 \otimes A_3 \otimes A_4 )
& = M_2 ( A_1, A_2 ) \otimes A_3 \otimes A_4
+(-1)^{{\rm deg} (A_1)} A_1 \otimes M_2 ( A_2, A_3 ) \otimes A_4 \,, \\
& \quad~ +(-1)^{{\rm deg} (A_1) +{\rm deg} (A_2)} A_1 \otimes A_2 \otimes M_2 ( A_3, A_4 ) \,, \\
& \quad \vdots
\end{split}
\end{equation}
and by
\begin{equation}
\begin{split}
{\bf M}_3 \, 1 & = 0 \,, \\
{\bf M}_3 \, A_1 & = 0 \,, \\
{\bf M}_3 \, ( A_1 \otimes A_2 ) & = 0 \,, \\
{\bf M}_3 \, ( A_1 \otimes A_2 \otimes A_3 )
& = M_3 ( A_1, A_2, A_3 ) \,, \\
{\bf M}_3 \, ( A_1 \otimes A_2 \otimes A_3 \otimes A_4 )
& = M_3 ( A_1, A_2, A_3 ) \otimes A_4
+(-1)^{{\rm deg} (A_1)} A_1 \otimes M_3 ( A_2, A_3, A_4 ) \,, \\
& \quad \vdots
\end{split}
\end{equation}
The $A_\infty$ relations can then be compactly written as
\begin{equation}
\begin{split}
& [ \, {\bf Q}, {\bf M}_2 \, ] \, ( A_1 \otimes A_2 ) = 0 \,, \\
& [ \, {\bf Q}, {\bf M}_3 \, ] \, ( A_1  \otimes A_2 \otimes A_3 )
+\frac{1}{2} \, [ \, {\bf M}_2, {\bf M}_2 \, ] \, ( A_1 \otimes A_2 \otimes A_3 ) = 0 \,.
\end{split}
\end{equation}
Consider an action written in terms of a set of degree-odd $n$-string products
$M_n ( A_1, A_2, \ldots , A_n )$
satisfying
\begin{equation}
\omega ( A_1, M_n ( A_2, A_3, \ldots , A_{n+1} ) )
= {}-(-1)^{{\rm deg} (A_1)} \, \omega ( M_n ( A_1, A_2, \ldots , A_n ), A_{n+1} )
\end{equation}
as
\begin{equation}
\begin{split}
S & = {}-\frac{1}{2} \, \omega ( \Psi, Q \Psi )
-\frac{1}{3} \, \omega ( \Psi, M_2 ( \Psi, \Psi ) )
-\frac{1}{4} \, \omega ( \Psi, M_3 ( \Psi, \Psi, \Psi ) ) + \ldots \\
& = {}-\sum_{n=1}^\infty \frac{1}{n+1} \,
\omega ( \Psi, M_n ( \, \underbrace{\Psi, \Psi, \ldots ,\Psi}_n \, ) ) \,,
\end{split}
\end{equation}
where
\begin{equation}
M_1 = Q \,.
\end{equation}
The $A_\infty$ relations can be expressed by introducing
${\bf M}_n$ acting on $T \mathcal{H}$
for the $n$-string product $M_n ( A_1, A_2, \ldots , A_n )$ as
\begin{equation}
[ \, {\bf M}, {\bf M} \, ] = 0 \,,
\end{equation}
where
\begin{equation}
{\bf M} = \sum_{n=1}^\infty {\bf M}_n
\end{equation}
with
\begin{equation}
{\bf M}_1 = {\bf Q} \,.
\end{equation}
The operators such as ${\bf M}_n$ are called {\it coderivations}.

We construct ${\bf M}$ satisfying the $A_\infty$ relations from the star product.
We introduce the coderivation ${\bf m}_2$ associated with $m_2$.
The associativity of the star product and the derivation property of $Q$ and $\eta$
with respect to the star product in~\eqref{m_2-properties}
can be stated in terms of ${\bf m}_2$ as
\begin{equation}
[ \, {\bf m}_2, {\bf m}_2 \, ] = 0 \,, \qquad
[ \, {\bf Q}, {\bf m}_2 \, ] = 0 \,, \qquad
[ \, {\bm \eta}, {\bf m}_2 \, ] = 0 \,.
\end{equation}

\subsection{Construction of the multi-string products}
\label{A_infinity-theory}

In~\cite{Erler:2013xta}
the NS sector of open superstring field theory
with the $A_\infty$ structure was constructed.
We introduce ${\bf M} (s)$ defined by
\begin{equation}
{\bf M} (s) = \sum_{n=0}^\infty s^n \, {\bf M}_{n+1} \,.
\end{equation}
The $A_\infty$ relations are
\begin{equation}
[ \, {\bf M} (s), {\bf M} (s) \, ] = 0 \,,
\end{equation}
and the condition that the string field $M_n ( \Psi, \Psi, \ldots, \Psi )$
is in the small Hilbert space can be stated as
\begin{equation}
[ \, {\bm \eta}, {\bf M} (s) \, ] = 0 \,.
\end{equation}
In the construction of~\cite{Erler:2013xta}, the coderivation ${\bf M} (s)$ is characterized
by the differential equation
\begin{equation}
\frac{d}{ds} \, {\bf M} (s)
= [ \, {\bf M} (s), {\bm \mu} (s) \, ]
\end{equation}
with the initial condition
\begin{equation}
{\bf M} (0) = {\bf Q} \,.
\end{equation}
The degree-even coderivation ${\bm \mu} (s)$ is arbitrary at the moment
except that the corresponding multi-string products $\mu_n$ have the following cyclic property:
\begin{equation}
\omega ( A_1, \mu_n ( A_2, A_3, \ldots , A_{n+1} ) )
= {}-\omega ( \, \mu_n ( A_1, A_2, \ldots , A_n ), A_{n+1} ) \,.
\end{equation}
We will discuss the conditions we impose on ${\bm \mu} (s)$ later.
The solution to the differential equation can be written as
\begin{equation}
{\bf M} (s) = {\bf G}^{-1} (s) \, {\bf Q} \, {\bf G} (s) \,,
\end{equation}
where ${\bf G} (s)$ is the path-ordered exponential given by
\begin{equation}
{\bf G} (s) = \mathcal{P} \exp \biggl[ \, \int_0^s ds' \, {\bm \mu} (s') \, \biggr]
\end{equation}
with the following ordering prescription:
\begin{equation}
\mathcal{P} \, [ \, {\bm \mu} (s_1) \, {\bm \mu} (s_2) \, ] = \biggl\{
\begin{array}{l}
{\bm \mu} (s_1) \, {\bm \mu} (s_2) \quad \text{for} \quad s_1 < s_2 \,, \\
{\bm \mu} (s_2) \, {\bm \mu} (s_1) \quad \text{for} \quad s_1 > s_2 \,.
\end{array}
\end{equation}
It obeys the differential equation
\begin{equation}
\frac{d}{ds} \, {\bf G} (s) = {\bf G} (s) \, {\bm \mu} (s)
\end{equation}
with the initial condition ${\bf G} (0) = 1$.
It then follows from the structure of ${\bf M} (s)$
that the $A_\infty$ relations are satisfied:
\begin{equation}
[ \, {\bf M} (s), {\bf M} (s) \, ]
= [ \, {\bf G}^{-1} (s) \, {\bf Q} \, {\bf G} (s), {\bf G}^{-1} (s) \, {\bf Q} \, {\bf G} (s) \, ]
= {\bf G}^{-1} (s) \, [ \, {\bf Q}, {\bf Q} \, ] \, {\bf G} (s) = 0 \,,
\end{equation}
where we used $[ \, {\bf Q}, {\bf Q} \, ] = 0$.

If $[ \, {\bm \eta}, {\bm \mu} (s) \, ] = 0$, the resulting theory
is related to the free theory by field redefinition.
To obtain a nontrivial interacting theory, we need
$[ \, {\bm \eta}, {\bm \mu} (s) \, ] \ne 0$,
while the condition $[ \, {\bm \eta}, {\bf M} (s) \, ] = 0$ is satisfied.
In the construction of~\cite{Erler:2013xta}, the coderivation ${\bm \mu} (s)$ is characterized by
\begin{equation}
[ \, {\bm \eta}, {\bm \mu} (s) \, ] = {\bf m} (s) \,,
\end{equation}
where ${\bf m} (s)$ is a degree-odd coderivation obeying the differential equation 
\begin{equation}
\frac{d}{ds} \, {\bf m} (s) = [ \, {\bf m} (s), {\bm \mu} (s) \, ]
\end{equation}
with the initial condition
\begin{equation}
{\bf m} (0 ) = {\bf m}_2 \,.
\end{equation}
The solution to the differential equation is given by
\begin{equation}
{\bf m} (s) = {\bf G}^{-1} (s) \, {\bf m}_2 \, {\bf G} (s) \,.
\label{m(s)-solution}
\end{equation}
Let us calculate $[ \, {\bm \eta}, {\bf G} (s) \, ]$ to show
that the condition $[ \, {\bm \eta}, {\bf M} (s) \, ] = 0$
is satisfied when we construct ${\bf M} (s)$ from ${\bm \mu} (s)$
characterized this way.
We have
\begin{equation}
[ \, {\bm \eta}, {\bf G} (s) \, ]
= \int_0^s ds' \, {\bf G} ( 0, s' ) \,
[ \, {\bm \eta}, {\bm \mu} (s') \, ] \, {\bf G} ( s', s )
= \int_0^s ds' \, {\bf G} ( 0, s' ) \, {\bf m} (s') \, {\bf G} ( s', s ) \,,
\end{equation}
where
\begin{equation}
{\bf G} (s_1, s_2)
= \mathcal{P} \exp \biggl[ \, \int_{s_1}^{s_2} ds' \, {\bm \mu} (s') \biggr] \,.
\end{equation}
It follows from~\eqref{m(s)-solution} that
\begin{equation}
{\bf G} ( 0, s' ) \, {\bf m} (s') = {\bf G} (s') \, {\bf m} (s')
= {\bf m}_2 \, {\bf G} (s') = {\bf m}_2 \, {\bf G} ( 0, s' ) \,,
\end{equation}
and we find
\begin{equation}
\int_0^s ds' \, {\bf G} ( 0, s' ) \, {\bf m} (s') \, {\bf G} ( s', s )
= \int_0^s ds' \, {\bf m}_2 \, {\bf G} ( 0, s' ) \, {\bf G} ( s', s )
= s \, {\bf m}_2 \, {\bf G} (s) \,.
\end{equation}
We thus obtain the following important relation:
\begin{equation}
[ \, {\bm \eta}, {\bf G} (s) \, ]
= s \, {\bf m}_2 \, {\bf G} (s) \,.
\label{[eta,G(s)]}
\end{equation}
Using this relation, we can show in the following way that the condition
$[ \, {\bm \eta}, {\bf M} (s) \, ] = 0$ is satisfied:
\begin{equation}
\begin{split}
[ \, {\bm \eta}, {\bf M} (s) \, ]
& = {}-{\bf G}^{-1} (s) \, [ \, {\bm \eta}, {\bf G} (s) \, ] \, {\bf G}^{-1} (s) \, {\bf Q} \, {\bf G} (s)
-{\bf G}^{-1} (s) \, {\bf Q} \, [ \, {\bm \eta}, {\bf G} (s) \, ] \\
& = {}-s \, {\bf G}^{-1} (s) \, {\bf m}_2 \, {\bf Q} \, {\bf G} (s)
-s \, {\bf G}^{-1} (s) \, {\bf Q} \, {\bf m}_2 \, {\bf G} (s)
= {}-s \, {\bf G}^{-1} (s) \, [ \, {\bf m}_2, {\bf Q} \, ] \, {\bf G} (s) = 0 \,,
\end{split}
\end{equation}
where we used $[ \, {\bf m}_2, {\bf Q} \, ] = 0$.

The path-ordered exponential ${\bf G} (s_1, s_2)$ is called a {\it cohomomorphism}.
A cohomomorphism ${\bf H}$ generates a field redefinition $H (\Psi)$
when it acts on a group-like element $1 / (1-\Psi)$,
which is defined for a degree-even state $\Psi$ by
\begin{equation}
\frac{1}{1-\Psi} = \sum_{n=0}^\infty \,
\underbrace{\, \Psi \otimes \Psi \otimes \ldots \otimes \Psi \,}_n
= 1 +\Psi +\Psi \otimes \Psi +\Psi \otimes \Psi \otimes \Psi \ldots \,,
\end{equation}
as follows:
\begin{equation}
H (\Psi) = \pi_1 \, {\bf H} \, \frac{1}{1-\Psi} \,,
\end{equation}
where $\pi_1$ is the projector to the one-string sector.
We only use a few identities regarding cohomomorphisms in this paper,
and they are summarized in appendix~\ref{cohomomorphism}.
See~\cite{Erler:2015small} for more detailed discussions.

\section{The relation to the Berkovits formulation}
\label{Berkovits-section}
\setcounter{equation}{0}

In this section we show that open superstring field theory
with the $A_\infty$ structure in the preceding section
is related to the Berkovits formulation
by partial gauge fixing and field redefinition.

\subsection{Transforming the action to the WZW-like form}
\label{WZW-like-form}

The action with the $A_\infty$ structure in the preceding section can be written as
\begin{equation}
\begin{split}
S & = {}-\frac{1}{2} \, \omega ( \Psi, Q \Psi )
-\frac{1}{3} \, \omega ( \Psi, M_2 ( \Psi, \Psi ) )
-\frac{1}{4} \, \omega ( \Psi, M_3 ( \Psi, \Psi, \Psi ) ) + \ldots \\
& = {}-\int_0^1 dt \, \biggl[ \, \omega ( \Psi, Q \, t \Psi )
+\omega ( \Psi, M_2 ( t \Psi, t \Psi ) )
+\omega ( \Psi, M_3 ( t \Psi, t \Psi, t \Psi ) ) + \ldots \, \biggr] \\
& = {}-\int_0^1 dt \, \sum_{n=1}^\infty
\omega ( \Psi, M_n ( \, \underbrace{t \Psi, t \Psi, \ldots ,t \Psi}_n \, ) )
= {}-\int_0^1 dt \, \omega \Bigl( \, \Psi, \pi_1 \, {\bf M} \, \frac{1}{1-t \Psi} \, \Bigr) \,,
\end{split}
\end{equation}
where ${\bf M} = {\bf M} (1)$.
The two important relations we derived in subsection~\ref{A_infinity-theory} are
\begin{equation}
{\bf M} = {\bf G}^{-1} \, {\bf Q} \, {\bf G}
\label{GM=QG}
\end{equation}
with
\begin{equation}
{\bf G} = {\bf G} (1)
= \mathcal{P} \exp \biggl[ \, \int_0^1 ds \, {\bm \mu} (s) \biggr] \,,
\end{equation}
and
\begin{equation}
[ \, {\bm \eta}, {\bf G} \, ]
= {\bf m}_2 \, {\bf G} \,,
\label{[eta,G]}
\end{equation}
which is~\eqref{[eta,G(s)]} with $s=1$.

We define the inner product $\omega_L$ in the large Hilbert space by
\begin{equation}
\omega_L \, ( \, \xi_0 A_1, A_2 ) = {}-\omega \, ( A_1, A_2 ) \,.
\end{equation}
It is related to the BPZ inner product in the large Hilbert space as
\begin{equation}
\omega_L \, ( \, A_1, A_2 ) = (-1)^{{\rm deg} (A_1)} \langle \, A_1, A_2 \, \rangle \,,
\end{equation}
and the inner product $\omega_L ( A_1, A_2 )$ is graded antisymmetric:
\begin{equation}
\omega_L \, ( A_1, A_2 )
= {}-(-1)^{{\rm deg} (A_1) \, {\rm deg} (A_2)} \, \omega_L \, ( A_2, A_1 ) \,.
\end{equation}
In terms of the inner product $\omega_L$, we have
\begin{equation}
S = \int_0^1 dt \, \omega_L \Bigl( \, \xi \Psi, \pi_1 {\bf M} \, \frac{1}{1 -t \Psi} \, \Bigr) \,.
\end{equation}
Since ${\bf M}$ consists of $Q$ and $\mu_n$,
we can use the cyclic property of $\mu_n$
to bring this action to the form
\begin{equation}
S = -\int_0^1 dt \sum_i \, \langle \, A_t^{(i)} (t) , Q A_\eta^{(i)} (t) \, \rangle
\end{equation}
with some $t$-dependent string fields $A_\eta^{(i)} (t)$ and $A_t^{(i)} (t)$.
It will turn out that the summation over $i$ is not necessary,
and the action will be brought to the form
\begin{equation}
S = -\int_0^1 dt \, \langle \, A_t (t) , Q A_\eta (t) \, \rangle \,.
\end{equation}
Let us first introduce the operator $\xi \partial_t$ as a one-string product
and denote the corresponding coderivation by ${\bm \xi}_t$.
We then write the action as
\begin{equation}
S = \int_0^1 dt \, \omega_L \Bigl( \,
\pi_1 \, {\bm \xi}_t  \, \frac{1}{1 -t \Psi}, \pi_1 {\bf M} \, \frac{1}{1 -t \Psi} \, \Bigr) \,.
\end{equation}
Using the identity~\eqref{G-field-redefinition} in appendix~\ref{cohomomorphism}, we have
\begin{equation}
\begin{split}
& \omega_L \Bigl( \,
\pi_1 \, {\bm \xi}_t  \, \frac{1}{1 -t \Psi}, \pi_1 {\bf M} \, \frac{1}{1 -t \Psi} \, \Bigr)
= \omega_L \Bigl( \,
\pi_1 \, {\bf G} \, {\bm \xi}_t  \, \frac{1}{1 -t \Psi},
\pi_1 {\bf G} \, {\bf M} \, \frac{1}{1 -t \Psi} \, \Bigr) \\
& = \omega_L \Bigl( \,
\pi_1 \, {\bf G} \, {\bm \xi}_t  \, \frac{1}{1 -t \Psi},
\pi_1 {\bf Q} \, {\bf G} \, \frac{1}{1 -t \Psi} \, \Bigr)
= \omega_L \Bigl( \,
\pi_1 \, {\bf G} \, {\bm \xi}_t  \, \frac{1}{1 -t \Psi},
Q \, \pi_1 {\bf G} \, \frac{1}{1 -t \Psi} \, \Bigr) \,,
\end{split}
\end{equation}
where we also used~\eqref{GM=QG}.
The action can therefore be written as
\begin{equation}
S = \int_0^1 dt \, \omega_L \Bigl( \,
\pi_1 {\bf G} \, {\bm \xi}_t  \, \frac{1}{1 -t \Psi},
Q \, \pi_1 {\bf G} \, \frac{1}{1 -t \Psi} \, \Bigr) \,,
\label{linear-t}
\end{equation}
and we identify $A_\eta (t)$ and $A_t (t)$ as
\begin{equation}
A_\eta (t) = \pi_1 \, {\bf G} \, \frac{1}{1 -t \Psi} \,, \quad
A_t (t) = \pi_1 \, {\bf G} \, {\bm \xi}_t \, \frac{1}{1 -t \Psi}
= \pi_1 \, {\bf G} \, \Bigl( \,
\frac{1}{1 -t \Psi} \otimes \xi \Psi \otimes \frac{1}{1 -t \Psi} \, \Bigr) \,.
\label{A_eta-A_t}
\end{equation}
In the context of the relation between cohomomorphisms and field redefinitions
mentioned at the end of section~\ref{A_infinity-section},
$A_\eta (t)$ is interpreted as the string field we obtain from $t \Psi$
by the field redefinition associated with the cohomomorphism ${\bf G}$.
This field redefinition, however, is not a map
to the small Hilbert space because $[ \, {\bm \eta}, {\bf G} \, ] \ne 0$.

An important property of the string field $A_\eta (t)$ is that it obeys the following equation
for any $t$:
\begin{equation}
\eta A_\eta (t) = A_\eta (t) A_\eta (t) \,.
\label{A_eta-relation}
\end{equation}
This can be shown as follows:
\begin{equation}
\begin{split}
\eta A_\eta (t) & = \pi_1 \, {\bm \eta} \, {\bf G} \, \frac{1}{1 -t \Psi}
= \pi_1 \, [ \, {\bm \eta}, {\bf G} \, ] \, \frac{1}{1 -t \Psi}
= \pi_1 \, {\bf m}_2 \, {\bf G} \, \frac{1}{1 -t \Psi} \\
& = \pi_1 \, {\bf m}_2 \, \frac{1}{1 -A_\eta (t)}
= A_\eta (t) A_\eta (t) \,,
\label{eta-A_eta-proof}
\end{split}
\end{equation}
where we used~\eqref{[eta,G]} and the identity~\eqref{cohomomorphism-identity}
in appendix~\ref{cohomomorphism}.
The relation~\eqref{A_eta-relation} is analogous to the equation of motion
$Q A +A^2 = 0$ in open bosonic string field theory
with the bosonic string field $A$.
In this case, however, the cohomology of $\eta$ is trivial,
and we can use $\xi$ as a homotopy operator satisfying $\{ \, \eta, \xi \, \} = 1$.
Therefore, the string field $A_\eta (t)$ satisfying~\eqref{A_eta-relation}
can be written in the pure-gauge form
\begin{equation}
A_\eta (t) = ( \eta e^{\xi \widetilde{\Psi} (t)} ) \, e^{-\xi \widetilde{\Psi} (t)}
\label{pure-gauge}
\end{equation}
with $\widetilde{\Psi} (t)$ in the small Hilbert space.

Furthermore,
since the string field $A_\eta (t)$ satisfies the relation~\eqref{A_eta-relation} for any $t$,
an infinitesimal change in $t$ has to be implemented by a gauge transformation
generated by $\eta$.
It is convenient to define the corresponding covariant derivative $D_\eta (t)$ by
\begin{equation}
D_\eta (t) \, \Phi = \eta \, \Phi -A_\eta (t) \, \Phi +(-1)^\Phi \, \Phi A_\eta (t) \,.
\end{equation}
It is nilpotent because of~\eqref{A_eta-relation}:
\begin{equation}
D_\eta (t)^2 = 0 \,.
\end{equation}
It acts as a derivation with respect to the star product,
\begin{equation}
D_\eta (t) \, ( \Phi_1 \Phi_2 )
= ( D_\eta (t) \, \Phi_1 ) \, \Phi_2 +(-1)^{\Phi_1} \Phi_1 \, ( D_\eta (t) \, \Phi_2 ) \,,
\end{equation}
and it is BPZ odd:
\begin{equation}
\langle \, \Phi_1, D_\eta (t) \, \Phi_2 \, \rangle
= {}-(-1)^{\Phi_1} \langle \, D_\eta (t) \, \Phi_1, \Phi_2 \, \rangle \,.
\end{equation}
Using the covariant derivative $D_\eta (t)$, $\partial_t A_\eta (t)$ can be written as
\begin{equation}
\partial_t A_\eta (t) = D_\eta (t) \, A_t (t)
= \eta A_t (t) -A_\eta (t) A_t (t) +A_t (t) A_\eta (t)  \,.
\label{partial_t-A_eta}
\end{equation}
As alluded by the notation, $A_t (t)$ in~\eqref{A_eta-A_t} satisfies this relation.
To prove this, let us calculate $\eta A_t (t)$. We find
\begin{align}
\eta A_t (t) & = \pi_1 \, {\bm \eta} \, {\bf G} \, {\bm \xi}_t \, \frac{1}{1 -t \Psi}
= \pi_1 \, [ \, {\bm \eta}, {\bf G} \, {\bm \xi}_t \, ] \, \frac{1}{1 -t \Psi}
= \pi_1 \, [ \, {\bm \eta}, {\bf G} \, ] \, {\bm \xi}_t \, \frac{1}{1 -t \Psi}
+\pi_1 \, {\bf G} \, [ \, {\bm \eta}, {\bm \xi}_t \, ] \, \frac{1}{1 -t \Psi} \nonumber \\
& = \pi_1 \, {\bf m}_2 \, {\bf G}  \, {\bm \xi}_t \, \frac{1}{1 -t \Psi}
+\pi_1 \, {\bf G} \, {\bm \partial}_t \, \frac{1}{1 -t \Psi} \,,
\end{align}
where ${\bm \partial}_t$ is the coderivation corresponding to $\partial_t$
as a one-string product.
We can use the identity~\eqref{cohomomorphism-coderivation-identity}
to transform the first term in the second line as
\begin{equation}
\begin{split}
\pi_1 \, {\bf m}_2 \, {\bf G}  \, {\bm \xi}_t \, \frac{1}{1 -t \Psi}
& = \pi_1 \, {\bf m}_2 \, \Bigl( \,
\frac{1}{1-A_\eta (t)} \otimes A_t (t) \otimes \frac{1}{1-A_\eta (t)} \, \Bigr) \\
& = A_\eta (t) A_t (t) -A_t (t) A_\eta (t) \,,
\end{split}
\end{equation}
while the second term is
\begin{equation}
\pi_1 \, {\bf G} \, {\bm \partial}_t \, \frac{1}{1 -t \Psi}
= \partial_t \, \pi_1 \, {\bf G} \, \frac{1}{1 -t \Psi} = \partial_t A_\eta (t) \,.
\end{equation}
We have thus shown~\eqref{partial_t-A_eta} with $A_t (t)$ in~\eqref{A_eta-A_t}.

Note that $A_t (t)$ satisfying~\eqref{partial_t-A_eta} for given $A_\eta (t)$ is not unique.
Suppose that $A_t^{(1)} (t)$ and $A_t^{(2)} (t)$ both satisfy~\eqref{partial_t-A_eta}.
Then the difference $\Delta A_t (t) = A_t^{(1)} (t) -A_t^{(2)} (t)$ is annihilated
by $D_\eta (t)$:
\begin{equation}
D_\eta (t) \, \Delta A_t (t) = 0 \,.
\label{D_eta-Delta}
\end{equation}

\subsection{Topological $t$-dependence}
\label{topological}

When we write the action in the form~\eqref{linear-t},
we can replace $t \Psi$ by $\Psi (t)$ with a general $t$-dependence
satisfying
\begin{equation}
\Psi (1) = \Psi \,, \qquad \Psi (0) = 0 \,.
\end{equation}
While the action is written in terms of $\Psi (t)$,
the $t$-dependence is topological and the action is a functional of $\Psi$.
This can be shown in the following way.

We first define $A_\eta (t)$ by
\begin{equation}
A_\eta (t) = \pi_1 \, {\bf G} \, \frac{1}{1-\Psi (t)} \,.
\end{equation}
This satisfies
\begin{equation}
\eta A_\eta (t) = A_\eta (t) \, A_\eta (t)
\end{equation}
for any $\Psi (t)$ because we can replace $t \Psi$ by $\Psi (t)$
in the proof~\eqref{eta-A_eta-proof}.
As in the case of the linear $t$-dependence,
an infinitesimal change in $t$ has to be implemented
by a gauge transformation as follows:
\begin{equation}
\partial_t A_\eta (t) = D_\eta (t) A_t (t)
= \eta A_t (t) -A_\eta (t) A_t (t) +A_t (t) A_\eta (t) \,.
\label{F_t-eta}
\end{equation}
We can show that $A_t (t)$ given by
\begin{equation}
A_t (t) = \pi_1 \, {\bf G} \, {\bm \xi}_t \, \frac{1}{1-\Psi (t)}
\end{equation}
satisfies~\eqref{F_t-eta}
because again we can replace $t \Psi$ by $\Psi (t)$
in the proof of the relation~\eqref{partial_t-A_eta}.

Similarly, under the variation $\delta \Psi (t)$, the change $\delta A_\eta (t)$
has to be implemented by a gauge transformation
with some gauge parameter $A_\delta (t)$ as follows:
\begin{equation}
\delta A_\eta (t) = D_\eta (t) A_\delta (t)
= \eta A_\delta (t) -A_\eta (t) A_\delta (t) +A_\delta (t) A_\eta (t) \,.
\label{F_delta-eta}
\end{equation}
We can think of a map from $\Psi (t)$ to $\xi \delta \Psi (t)$
as an action of a one-string product, and we denote the corresponding coderivation
by ${\bm \xi}_\delta$.
We can then show that $A_\delta (t)$ given by
\begin{equation}
A_\delta (t) = \pi_1 \, {\bf G} \, {\bm \xi}_\delta \, \frac{1}{1-\Psi (t)}
\end{equation}
satisfies~\eqref{F_delta-eta} as in the proof of the relation~\eqref{partial_t-A_eta}.

We will also need a relation between $\delta A_t (t)$ and $\partial_t A_\delta (t)$
which follows from the equation 
$\delta \, \partial_t A_\eta (t) = \partial_t \, \delta A_\eta (t)$.
Starting with~\eqref{F_t-eta}, we find
\begin{equation}
\begin{split}
\delta \, \partial_t A_\eta (t)
& = \eta \, \delta A_t (t) -\delta A_\eta (t) A_t (t) -A_\eta (t) \delta A_t (t)
+\delta A_t (t) A_\eta (t) +A_t (t) \delta A_\eta (t) \\
& = D_\eta (t) \, \delta A_t (t) -\delta A_\eta (t) A_t (t) +A_t (t) \delta A_\eta (t) \\
& = D_\eta (t) \, \delta A_t (t) -( D_\eta (t) A_\delta (t) ) \, A_t (t) +A_t (t) \, ( D_\eta (t) A_\delta (t) ) \,.
\end{split}
\end{equation}
Starting with~\eqref{F_delta-eta}, we find
\begin{equation}
\begin{split}
\partial_t \, \delta A_\eta (t)
& = \eta \, \partial_t A_\delta (t) -\partial_t A_\eta (t) A_\delta (t) -A_\eta (t) \partial_t A_\delta (t)
+\partial_t A_\delta (t) A_\eta (t) +A_\delta (t) \partial_t A_\eta (t) \\
& = D_\eta (t) \, \partial_t A_\delta (t)
-\partial_t A_\eta (t) A_\delta (t) +A_\delta (t) \partial_t A_\eta (t) \\
& = D_\eta (t) \, \partial_t A_\delta (t)
-( D_\eta (t) A_t (t) ) \, A_\delta (t) +A_\delta (t) \, ( D_\eta A_t (t) ) \,.
\end{split}
\end{equation}
We therefore have
\begin{equation}
D_\eta (t) F_{\delta t} (t) = 0 \,,
\end{equation}
where
\begin{equation}
F_{\delta t} (t) = \delta A_t (t) -\partial_t A_\delta (t)
-A_\delta (t) A_t (t) +A_t (t) A_\delta (t) \,.
\label{F_delta-t}
\end{equation}

Let us now consider the variation of the action.
The action is given by
\begin{equation}
S = -\int_0^1 dt \, \langle \, A_t (t), Q A_\eta (t) \, \rangle \,.
\end{equation}
Under the variation $\delta \Psi (t)$, we have
\begin{equation}
\delta \, \langle \, A_t (t), Q A_\eta (t) \, \rangle
= \langle \, \delta A_t (t), Q A_\eta (t) \, \rangle
+\langle \, A_t (t), Q \, \delta A_\eta (t) \, \rangle \,.
\label{integrand-variation}
\end{equation}
Let us rewrite the first term on the right-hand side
using $F_{\delta t} (t)$ in~\eqref{F_delta-t}:
\begin{equation}
\begin{split}
\langle \, \delta A_t (t), Q A_\eta (t) \, \rangle
& = \langle \, \partial_t A_\delta (t), Q A_\eta (t) \, \rangle
+\langle \, A_\delta (t) A_t (t) -A_t (t) A_\delta (t), Q A_\eta (t) \, \rangle \\
& \quad~ +\langle \, F_{\delta t} (t), Q A_\eta (t) \, \rangle \,.
\end{split}
\end{equation}
The last term on the right-hand side actually vanishes 
because $Q A_\eta (t)$ can be written in the form
\begin{equation}
Q A_\eta (t) = {}-D_\eta (t) \, A_Q (t) \,.
\label{Q-A_eta}
\end{equation}
This can be shown as follows.
First, the coderivation ${\bf M}$ satisfies
the condition $[ \, {\bm \eta}, {\bf M} \, ] = 0$ so that
it can be written as
\begin{equation}
{\bf M} = {}-[ \, {\bm \eta}, {\bm \xi}_Q \, ]
\end{equation}
with some coderivation ${\bm \xi}_Q$.
We then have
\begin{equation}
\begin{split}
& Q A_\eta (t)
= \pi_1\, {\bf Q} \, {\bf G} \, \frac{1}{1-\Psi (t)}
= \pi_1\, {\bf G} \, {\bf M} \, \frac{1}{1-\Psi (t)}
= {}-\pi_1\, {\bf G} \, [ \, {\bm \eta}, {\bm \xi}_Q \, ] \, \frac{1}{1-\Psi (t)} \\
& = {}-\pi_1 {\bf G} \, {\bm \eta} \, {\bm \xi}_Q \, \frac{1}{1-\Psi (t)}
= {}-\pi_1 \, {\bm \eta} \, {\bf G} \, {\bm \xi}_Q \, \frac{1}{1-\Psi (t)}
+\pi_1 \, {\bf m}_2 \, {\bf G} \, {\bm \xi}_Q \, \frac{1}{1-\Psi (t)} \\
& = {}-\eta A_Q (t) +A_\eta (t) A_Q (t) +A_Q (t) A_\eta (t) = {}-D_\eta (t) \, A_Q (t)
\end{split}
\end{equation}
with
\begin{equation}
A_Q (t) = \pi_1 \, {\bf G} \, {\bm \xi}_Q \, \frac{1}{1-\Psi (t)} \,.
\end{equation}
In appendix~\ref{up-to-quartic},
we explicitly write $Q A_\eta (t)$ in this form to a few orders in $\Psi$
for demonstration.
Since $D_\eta (t) F_{\delta t} (t) = 0$, we find
\begin{equation}
\langle \, F_{\delta t} (t), Q A_\eta (t) \, \rangle
= {}-\langle \, F_{\delta t} (t), D_\eta (t) A_Q (t) \, \rangle
= \langle \, D_\eta (t) F_{\delta t} (t), A_Q (t) \, \rangle = 0 \,.
\end{equation}
For the second term on the right-hand side of~\eqref{integrand-variation},
we use~\eqref{F_delta-eta} and~\eqref{F_t-eta} to find
\begin{equation}
\begin{split}
& \langle \, A_t (t), Q \, \delta A_\eta (t) \, \rangle
= \langle \, A_t (t), Q \, \eta A_\delta (t) \, \rangle
-\langle \, A_t (t), Q \, ( \, A_\eta (t) A_\delta (t) -A_\delta (t) A_\eta (t) \, ) \, \rangle \\
& = \langle \, \eta A_t (t), Q A_\delta (t) \, \rangle
+\langle \, Q A_t (t), A_\eta (t) A_\delta (t) -A_\delta (t) A_\eta (t) \, \rangle \\
& = \langle \, \partial_t A_\eta (t), Q A_\delta (t) \, \rangle
+\langle \, A_\eta (t) A_t (t) -A_t (t) A_\eta (t), Q A_\delta (t) \, \rangle \\
& \quad~ +\langle \, Q A_t (t), A_\eta (t) A_\delta (t) -A_\delta (t) A_\eta (t) \, \rangle \\
& = \langle \, A_\delta (t), Q \, \partial_t A_\eta (t) \, \rangle
-\langle \, Q A_\delta (t), A_\eta (t) A_t (t) -A_t (t) A_\eta (t) \, \rangle \\
& \quad~ +\langle \, Q A_t (t), A_\eta (t) A_\delta (t) -A_\delta (t) A_\eta (t) \, \rangle \,.
\end{split}
\end{equation}
Since
\begin{equation}
\begin{split}
& \langle \, A_\delta (t) A_t (t) -A_t (t) A_\delta (t), Q A_\eta (t) \, \rangle
= {}-\langle \, Q \, ( \, A_\delta (t) A_t (t) -A_t (t) A_\delta (t) \, ), A_\eta (t) \, \rangle \\
& = {}-\langle \, Q A_\delta (t), A_t (t) A_\eta (t) -A_\eta (t) A_t (t) \, \rangle
+\langle \, Q A_t (t), A_\delta (t) A_\eta (t) -A_\eta (t) A_\delta (t) \, \rangle \,,
\end{split}
\end{equation}
we find
\begin{equation}
\delta \, \langle \, A_t (t), Q A_\eta (t) \, \rangle
= \langle \, \partial_t A_\delta (t), Q A_\eta (t) \, \rangle
+\langle \, A_\delta (t), Q \, \partial_t A_\eta (t) \, \rangle
= \partial_t \, \langle \, A_\delta (t), Q A_\eta (t) \, \rangle \,.
\end{equation}
This shows that the $t$-dependence of the action is topological:
\begin{equation}
\delta S = -\int_0^1 dt \, \partial_t \, \langle \, A_\delta (t), Q A_\eta (t) \, \rangle
= {}-\langle \, A_\delta, Q A_\eta \, \rangle \,,
\end{equation}
where
\begin{equation}
A_\eta = A_\eta (1) = \pi_1 \, {\bf G} \, \frac{1}{1-\Psi} \,, \quad
A_\delta = A_\delta (1)
= \pi_1 \, {\bf G} \, {\bm \xi}_\delta \, \frac{1}{1-\Psi}
= \pi_1 \, {\bf G} \, \Bigl( \,
\frac{1}{1-\Psi} \otimes \xi \delta \Psi \otimes \frac{1}{1-\Psi} \, \Bigr) \,.
\end{equation}

Let us verify that this form of the variation of the action
coincides with the original form we started with.
The variation of the action can be written as
\begin{equation}
\begin{split}
& \delta S = {}-\langle \, A_\delta, Q A_\eta \, \rangle
= \omega_L \Bigl( \, \pi_1 \, {\bf G} \, {\bm \xi}_\delta \, \frac{1}{1-\Psi},
Q \, \pi_1 \, {\bf G} \, \frac{1}{1-\Psi} \, \Bigr) \\
& = \omega_L \Bigl( \, \pi_1 \, {\bf G} \, {\bm \xi}_\delta \, \frac{1}{1-\Psi},
\pi_1 \, {\bf Q} \, {\bf G} \, \frac{1}{1-\Psi} \, \Bigr)
= \omega_L \Bigl( \, \pi_1 \, {\bf G} \, {\bm \xi}_\delta \, \frac{1}{1-\Psi},
\pi_1 \, {\bf G} \, {\bf M} \, \frac{1}{1-\Psi} \, \Bigr) \,.
\end{split}
\end{equation}
Using the identity~\eqref{G-field-redefinition}, we have
\begin{equation}
\begin{split}
\delta S & = \omega_L \Bigl( \, \pi_1 \, {\bm \xi}_\delta \, \frac{1}{1-\Psi},
\pi_1 \, {\bf M} \, \frac{1}{1-\Psi} \, \Bigr) \\
& = \omega_L \Bigl( \, \xi \, \delta \Psi, \pi_1 \, {\bf M} \, \frac{1}{1-\Psi} \, \Bigr)
= {}-\omega \Bigl( \, \delta \Psi, \pi_1 \, {\bf M} \, \frac{1}{1-\Psi} \, \Bigr) \,.
\end{split}
\end{equation}
The equation of motion is thus given by
\begin{equation}
\pi_1 \, {\bf M} \, \frac{1}{1-\Psi} = 0 \,,
\end{equation}
and the action is invariant under the gauge transformation given by
\begin{equation}
\delta_\Lambda \Psi = \pi_1 \, {\bf M} \, \Bigl( \,
\frac{1}{1-\Psi} \otimes \Lambda \otimes \frac{1}{1-\Psi} \, \Bigr) \,.
\end{equation}

\subsection{Field redefinition}
\label{field-redefinition}

To summarize, we have shown that the action constructed in~\cite{Erler:2013xta}
with the $A_\infty$ structure can be written in the form
\begin{equation}
S = -\int_0^1 dt \, \langle \, A_t (t), Q A_\eta (t) \, \rangle \,,
\label{action-Psi}
\end{equation}
where 
\begin{equation}
A_\eta (t) = \pi_1 {\bf G} \, \frac{1}{1-\Psi (t)} \,, \quad
A_t (t) = \pi_1 {\bf G} \, {\bm \xi}_t \frac{1}{1-\Psi (t)}
= \pi_1 {\bf G} \,
\Bigl( \, \frac{1}{1-\Psi (t)} \otimes \xi \partial_t \Psi (t) \otimes \frac{1}{1-\Psi (t)} \, \Bigr) \,,
\label{A_eta-A_t-Psi}
\end{equation}
and $A_\eta (t)$ and $A_t (t)$ satisfy
\begin{align}
\eta A_\eta (t) & = A_\eta (t) A_\eta (t) \,,
\label{eta-A_eta-again} \\
\partial_t A_\eta (t) & = D_\eta (t) \, A_t (t)
= \eta A_t (t) -A_\eta (t) A_t (t) +A_t (t) A_\eta (t) \,.
\label{partial_t-A_eta-again}
\end{align}
We show in this subsection that this action in terms of $\Psi$ is related
to the action in terms of $\widetilde{\Psi}$
which is obtained from the Berkovits formulation by the partial gauge fixing~\cite{Iimori:2013kha},
\begin{equation}
S = -\int_0^1 dt \, \langle \, \widetilde{A}_t (t), Q \widetilde{A}_\eta (t) \, \rangle
\end{equation}
with
\begin{equation}
\widetilde{A}_\eta (t) = ( \eta \, e^{\xi \widetilde{\Psi} (t)} ) \, e^{-\xi \widetilde{\Psi} (t)} \,, \quad
\widetilde{A}_t (t) = ( \partial_t \, e^{\xi \widetilde{\Psi} (t)} ) \, e^{-\xi \widetilde{\Psi} (t)} \,,
\end{equation}
by the field redefinition determined from
\begin{equation}
A_\eta (t) = \widetilde{A}_\eta (t) \,.
\label{A_eta=tilde-A_eta}
\end{equation}

In terms of the string fields $\Psi (t)$ and $\widetilde{\Psi} (t)$,
the relation~\eqref{A_eta=tilde-A_eta} is written as
\begin{equation}
\pi_1 {\bf G} \, \frac{1}{1-\Psi (t)}
= ( \eta \, e^{\xi \widetilde{\Psi} (t)} ) \, e^{-\xi \widetilde{\Psi} (t)} \,.
\label{field-redefinition-t}
\end{equation}
Using this relation between $\Psi (t)$ and $\widetilde{\Psi} (t)$,
we can write $A_t (t)$ in terms of $\widetilde{\Psi} (t)$.
The string fields $A_\eta (t)$ and $A_t (t)$ written in terms of $\widetilde{\Psi} (t)$
satisfy~\eqref{partial_t-A_eta-again}.
On the other hand, $\widetilde{A}_t (t)$ also satisfies
\begin{equation}
\partial_t A_\eta (t) = D_\eta (t) \, \widetilde{A}_t (t)
= \eta \widetilde{A}_t (t) -A_\eta (t) \widetilde{A}_t (t) +\widetilde{A}_t (t) A_\eta (t)
\end{equation}
because of the relation~\eqref{A_eta=tilde-A_eta}.
As we explained in~\eqref{D_eta-Delta}, 
the difference $\Delta A_t (t) = A_t (t) -\widetilde{A}_t (t)$
is therefore annihilated by~$D_\eta (t)$:
\begin{equation}
D_\eta (t) \, \Delta A_t (t) = 0 \,.
\end{equation}
It then follows that the difference does not contribute to the action because
\begin{equation}
\langle \, \Delta A_t (t), Q A_\eta (t) \, \rangle
= {}-\langle \, \Delta A_t (t), D_\eta (t) A_Q (t) \, \rangle
= \langle \, D_\eta (t) \, \Delta A_t (t), A_Q (t) \, \rangle = 0 \,,
\end{equation}
where we used~\eqref{Q-A_eta}.
We have thus shown that the action~\eqref{action-Psi}
written in terms of $\Psi (t)$ is mapped to
\begin{equation}
S = -\int_0^1 dt \, \langle \, ( \partial_t \, e^{\xi \widetilde{\Psi} (t)} ) \, e^{-\xi \widetilde{\Psi} (t)},
Q \, ( \, ( \eta \, e^{\xi \widetilde{\Psi} (t)} ) \, e^{-\xi \widetilde{\Psi} (t)} \, ) \, \rangle
\label{action-Psi-tilde}
\end{equation}
by the field redefinition determined from~\eqref{field-redefinition-t}.
Since the $t$-dependence is topological in both actions,
the action~\eqref{action-Psi} is a functional of $\Psi$ and
the action~\eqref{action-Psi-tilde} is a functional of $\widetilde{\Psi}$.
The string fields $\Psi$ and $\widetilde{\Psi}$ of the two theories are related by
\begin{equation}
\pi_1 \, {\bf G} \, \frac{1}{1-\Psi}
= ( \eta \, e^{\xi \widetilde{\Psi}} ) \, e^{-\xi \widetilde{\Psi}} \,.
\end{equation}

\subsection{Embedding to the large Hilbert space}
\label{embedding}

The theory in terms of $\widetilde{\Psi} (t)$ is obtained from the Berkovits formulation
in terms of $\widetilde{\Phi} (t)$ by the partial gauge fixing.
The string fields of the two theories are related by
\begin{equation}
\widetilde{\Phi} (t) = \xi \widetilde{\Psi} (t) \,.
\end{equation}
In fact, the theory in terms of $\Psi (t)$ also allows a description
in terms of a string field $\Phi (t)$ in the large Hilbert space,
where the string fields of the two theories are related by
\begin{equation}
\Phi (t) = \xi \Psi (t) \,.
\end{equation}
By replacing $\Psi (t)$ by $\eta \Phi (t)$
and $\xi \partial_t \Psi (t)$ by $\partial_t \Phi (t)$ in~\eqref{A_eta-A_t-Psi},
we have
\begin{equation}
A_\eta (t) = \pi_1 {\bf G} \, \frac{1}{1-\eta \, \Phi (t)} \,, \quad
A_t (t) = \pi_1 {\bf G} \, \Bigl( \,
\frac{1}{1-\eta \, \Phi (t)} \otimes \partial_t \Phi (t) \otimes \frac{1}{1-\eta \, \Phi (t)} \, \Bigr) \,.
\end{equation}
Apparently, the action in terms of $\Phi (t)$ does not contain $\xi$,
but we use $\xi$ to realize ${\bf G}$.
While the operator $\xi$ does not have simple transformation properties
under conformal transformations,
the relations~\eqref{eta-A_eta-again} and~\eqref{partial_t-A_eta-again}
are satisfied
and they guarantee that the action has gauge invariances in the large Hilbert space.

The variation of the action is given by
\begin{equation}
\delta S = \omega_L \Bigl( \, \delta \Phi, \pi_1 \, {\bf M} \, \frac{1}{1-\eta \Phi} \, \Bigr) \,.
\end{equation}
The equation of motion is
\begin{equation}
\pi_1 \, {\bf M} \, \frac{1}{1-\eta \Phi} = 0 \,,
\end{equation}
and the action is invariant under the gauge transformation given by
\begin{equation}
\delta \Phi = \pi_1 \, {\bf M} \,
\Bigl( \, \frac{1}{1-\eta \Phi} \otimes \Lambda \otimes \frac{1}{1-\eta \Phi} \, \Bigr)
+\eta \Omega \,,
\end{equation}
where $\Lambda$ and $\Omega$ are gauge parameters.
Note that the gauge transformation generated by $\eta$ is linear
even in the interacting theory.
If we choose $\Phi (t) = t \Phi$, the action can be written as
\begin{equation}
\begin{split}
S & = \frac{1}{2} \, \omega_L ( \Phi, Q \eta \Phi )
+\sum_{n=2}^\infty \frac{1}{n+1} \,
\omega_L ( \Phi, M_n ( \, \underbrace{\eta \Phi, \eta \Phi, \ldots , \eta \Phi}_n \, ) ) \\
& = \frac{1}{2} \, \omega_L ( \Phi, Q \eta \Phi )
+\frac{1}{3} \, \omega_L ( \Phi, M_2 ( \eta \Phi, \eta \Phi ) )
+\frac{1}{4} \, \omega_L ( \Phi, M_3 ( \eta \Phi, \eta \Phi, \eta \Phi ) ) + \ldots \,.
\end{split}
\label{Phi-linear-t}
\end{equation}
For any gauge-invariant theory in terms of $\Psi$ in the small Hilbert space,
we can trivially embed it to the theory with $\Phi$ in the large Hilbert space
by replacing $\Psi$ with $\eta \Phi$.
While the form~\eqref{Phi-linear-t} of the action can be directly obtained
by this procedure, an important point in our discussion is that the actions
written in terms of $\Psi (t)$ and $\Phi (t)$ both take the WZW-like form.

\section{Conclusions and discussion}
\label{conclusions-discussion}
\setcounter{equation}{0}

We have considered four theories in this paper.
(See figure~\ref{relation-figure} in the introduction.)
In the Berkovits formulation
the string field $\widetilde{\Phi}$ is in the large Hilbert space,
and the theory in terms of $\widetilde{\Psi}$
in the small Hilbert space is obtained from the Berkovits formulation
by the partial gauge fixing~\cite{Iimori:2013kha}.
We have also seen that the theory with the $A_\infty$ structure
constructed in~\cite{Erler:2013xta}
written in terms of $\Psi$ in the small Hilbert space
can be obtained from a theory
written in terms of $\Phi$ in the large Hilbert space
by the partial gauge fixing.
The actions of these four theories can be written
in the following common form:
\begin{equation}
S = -\int_0^1 dt \, \langle \, A_t (t), Q A_\eta (t) \, \rangle \,,
\end{equation}
where 
and $A_\eta (t)$ and $A_t (t)$ satisfy
\begin{align}
\eta A_\eta (t) & = A_\eta (t) A_\eta (t) \,,
\label{eta-A_eta-once-again} \\
\partial_t A_\eta (t) & = \eta A_t (t) -A_\eta (t) A_t (t) +A_t (t) A_\eta (t) \,.
\label{partial_t-A_eta--once-again}
\end{align}
The string fields $A_\eta (t)$ and $A_t (t)$ are parameterized
by $\Phi (t)$, $\Psi (t)$, $\widetilde{\Phi} (t)$, and $\widetilde{\Psi} (t)$ as
\begin{align}
A_\eta (t) & = \pi_1 {\bf G} \, \frac{1}{1-\eta \, \Phi (t)} \,, \quad
A_t (t) = \pi_1 {\bf G} \, \Bigl( \,
\frac{1}{1-\eta \, \Phi (t)} \otimes \partial_t \Phi (t) \otimes \frac{1}{1-\eta \, \Phi (t)} \, \Bigr) \,, \\
A_\eta (t) & = \pi_1 {\bf G} \, \frac{1}{1-\Psi (t)} \,, \quad
A_t (t) = \pi_1 {\bf G} \,
\Bigl( \, \frac{1}{1-\Psi (t)} \otimes \xi \partial_t \Psi (t) \otimes \frac{1}{1-\Psi (t)} \, \Bigr) \,, \\
A_\eta (t) & = ( \eta \, e^{\widetilde{\Phi} (t)} ) \, e^{-\widetilde{\Phi} (t)} \,, \quad
A_t (t) = ( \partial_t \, e^{\widetilde{\Phi} (t)} ) \, e^{-\widetilde{\Phi} (t)} \,, \\
A_\eta (t) & = ( \eta \, e^{\xi \widetilde{\Psi} (t)} ) \, e^{-\xi \widetilde{\Psi} (t)} \,, \quad
A_t (t) = ( \partial_t \, e^{\xi \widetilde{\Psi} (t)} ) \, e^{-\xi \widetilde{\Psi} (t)} \,.
\end{align}
These string fields are related by the partial gauge fixing as follows:
\begin{equation}
\Phi (t) = \xi \Psi (t) \,, \qquad \widetilde{\Phi} (t) = \xi \widetilde{\Psi} (t) \,.
\end{equation}
The two string fields $\Psi (t)$ and $\widetilde{\Psi} (t)$ in the small Hilbert space are related by
\begin{equation}
\pi_1 {\bf G} \, \frac{1}{1-\Psi (t)}
= ( \eta \, e^{\xi \widetilde{\Psi} (t)} ) \, e^{-\xi \widetilde{\Psi} (t)} \,.
\end{equation}
The $t$ dependence in these actions is topological,
and the actions are functionals of $\Psi$, $\widetilde{\Psi}$, $\Phi$, and $\widetilde{\Phi}$,
which are values of $\Psi (t)$, $\widetilde{\Psi} (t)$, $\Phi (t)$, and $\widetilde{\Phi} (t)$,
respectively,  at $t=1$.
The field redefinition between $\Psi$ and $\widetilde{\Psi}$ is determined by
\begin{equation}
\pi_1 {\bf G} \, \frac{1}{1-\Psi}
= ( \eta \, e^{\xi \widetilde{\Psi}} ) \, e^{-\xi \widetilde{\Psi}} \,.
\label{Psi-tilde-Psi}
\end{equation}

The relation between $\Phi$ and $\widetilde{\Phi}$ in the large Hilbert space
is essentially determined by the relation between $\Psi$ and $\widetilde{\Psi}$
in the small Hilbert space, but there are many interesting aspects to be explored
for the embedding of the theory with the $A_\infty$ structure to the large Hilbert space
and its relation to the Berkovits formulation.
We hope to continue research in this direction.

Another important direction to extend the results in this paper
will be to consider general $A_\infty$ structures
with nonassociative two-string products,
as discussed in~\cite{Erler:2014eba}.
Once we succeed in this generalization,
we expect that the generalization to heterotic string field theory
would be straightforward,
and we hope to further extend the results in this paper
to type II superstring field theory.\footnote{
See~\cite{Jurco:2013qra, Matsunaga:2013mba, Matsunaga:2014wpa}
for recent discussions on closed superstring field theory.
}

Use of the large Hilbert space of the superconformal ghost sector
has been an important ingredient in formulating superstring field theory,
and we believe that we have gained insight into its relation to the $A_\infty$ structure.
Another crucial aspect in formulating superstring field theory
would be its relation to the supermoduli space of super-Riemann surfaces~\cite{Ohmori-Okawa},
and when we combine all the insight from the large Hilbert space,
the $A_\infty$ structure, and the supermoduli space of super-Riemann surfaces,
we expect to have more fundamental understanding
towards completion of formulating superstring field theory.

\bigskip
\noindent
{\bf \large Acknowledgments}

\medskip
The work of T.E. was supported in part
by the DFG Transregional Collaborative Research Centre TRR 33
and the DFG cluster of excellence Origin and Structure of the Universe.
The work of Y.O. was supported in part
by Grant-in-Aid for Scientific Research~(B) No.~25287049
and Grant-in-Aid for Scientific Research~(C) No.~24540254
from the Japan Society for the Promotion of Science (JSPS).

\appendix

\section{Coderivations and cohomomorphisms}
\label{cohomomorphism}
\setcounter{equation}{0}

In this appendix we summarize identities regarding cohomomorphisms
used in this paper. See~\cite{Erler:2015small} for more detailed discussions.

For a degree-even state $\Psi$,
we define a group-like element $1/(1-\Psi)$ by
\begin{equation}
\frac{1}{1-\Psi} = \sum_{n=0}^\infty \,
\underbrace{\, \Psi \otimes \Psi \otimes \ldots \otimes \Psi \,}_n
= 1 +\Psi +\Psi \otimes \Psi +\Psi \otimes \Psi \otimes \Psi \ldots \,.
\end{equation}
The action of a coderivation ${\bf C}$ on $1/(1-\Psi)$ is given by
\begin{equation}
{\bf C} \, \frac{1}{1-\Psi}
= \frac{1}{1-\Psi} \otimes \Bigl( \, \pi_1 \, {\bf C} \frac{1}{1-\Psi} \, \Bigr) \otimes \frac{1}{1-\Psi} \,.
\end{equation}
Consider the cohomomorphism ${\bf H}$ given by
\begin{equation}
{\bf H} = {\bf G} (s_1, s_2)
= \mathcal{P} \exp \biggl[ \, \int_{s_1}^{s_2} ds' \, {\bm \mu} (s') \biggr] \,.
\end{equation}
It generates a field redefinition $H (\Psi)$
when it acts on $1/(1-\Psi)$ as follows:
\begin{equation}
H (\Psi) = \pi_1 \, {\bf H} \, \frac{1}{1-\Psi} \,.
\end{equation}
In the rest of this appendix, we prove the following identities:
\begin{align}
\label{cohomomorphism-identity}
{\bf H} \, \frac{1}{1-\Psi} & = \frac{1}{1-H(\Psi)} \,, \\
\label{cohomomorphism-coderivation-identity}
{\bf H} \, {\bf C} \, \frac{1}{1-\Psi}
& = \frac{1}{1-H(\Psi)}
\otimes \Bigl( \, \pi_1 \, {\bf H} \, {\bf C} \, \frac{1}{1-\Psi} \, \Bigr)
\otimes \frac{1}{1-H(\Psi)} \,, \\
\label{H-field-redefinition}
\omega_L \Bigl( \, \pi_1 {\bf C}_1 \, \frac{1}{1 -\Psi},
\pi_1 {\bf C}_2 \, \frac{1}{1 -\Psi} \, \Bigr)
& = \omega_L \Bigl( \, \pi_1 {\bf H} \, {\bf C}_1 \, \frac{1}{1 -\Psi},
\pi_1 {\bf H} \, {\bf C}_2 \, \frac{1}{1 -\Psi} \, \Bigr) \,,
\end{align}
where ${\bf C}$, ${\bf C}_1$, and ${\bf C}_2$ are arbitrary coderivations.

Let us begin with~\eqref{cohomomorphism-identity}.
This is trivially satisfied when $s_1 = s_2$.
We prove~\eqref{cohomomorphism-identity} by showing
\begin{equation}
{\bf G} (s-\Delta s, s_2) \, \frac{1}{1-\Psi} = \frac{1}{1-G(s-\Delta s; \Psi)}
+O(\Delta s^2)
\label{for-identity-1}
\end{equation}
with
\begin{equation}
G (s; \Psi) = \pi_1 \, {\bf G} (s, s_2) \, \frac{1}{1-\Psi}
\end{equation}
when
\begin{equation}
{\bf G} (s, s_2) \, \frac{1}{1-\Psi} = \frac{1}{1-G(s; \Psi)}
\label{assumption-1}
\end{equation}
is satisfied.
From the definition of the path-ordered exponential, ${\bf G} (s-\Delta s, s_2)$ is given by
\begin{equation}
{\bf G} (s-\Delta s, s_2) = {\bf G} (s, s_2) +\Delta s \, {\bm \mu} (s) \, {\bf G} (s, s_2)
+O(\Delta s^2) \,.
\end{equation}
Assuming~\eqref{assumption-1},
the left-hand side of~\eqref{for-identity-1} is then
\begin{equation}
\begin{split}
& {\bf G} (s-\Delta s, s_2) \, \frac{1}{1-\Psi}
= {\bf G} (s, s_2) \, \frac{1}{1-\Psi}
+\Delta s \, {\bm \mu} (s) \, {\bf G} (s, s_2) \, \frac{1}{1-\Psi} +O(\Delta s^2) \\
& = \frac{1}{1-G (s; \Psi)}
+\Delta s \, {\bm \mu} (s) \, \frac{1}{1-G (s; \Psi)} +O(\Delta s^2) \\
& = \frac{1}{1-G(s; \Psi)}
+\Delta s \, \frac{1}{1-G(s; \Psi)}
\otimes \Bigl( \, \pi_1 \, {\bm \mu} (s) \, \frac{1}{1-G (s; \Psi)} \, \Bigr)
\otimes \frac{1}{1-G(s; \Psi)} +O(\Delta s^2) \,.
\end{split}
\end{equation}
On the other hand, $G(s-\Delta s; \Psi)$ is given by
\begin{equation}
\begin{split}
G(s-\Delta s; \Psi)
& = \pi_1 \, {\bf G} (s, s_2) \, \frac{1}{1-\Psi}
+\Delta s \, \pi_1 \, {\bm \mu} (s) \, {\bf G} (s, s_2) \, \frac{1}{1-\Psi}
+O(\Delta s^2) \\
& = G (s; \Psi)
+\Delta s \, \pi_1 \, {\bm \mu} (s) \, \frac{1}{1-G (s; \Psi)} +O(\Delta s^2) \,,
\end{split}
\end{equation}
and $1/(1-G(s-\Delta s; \Psi) )$ is
\begin{equation}
\begin{split}
& \frac{1}{1-G(s-\Delta s; \Psi)} \\
& = \frac{1}{1-G(s; \Psi)}
+\Delta s \, \frac{1}{1-G(s; \Psi)}
\otimes \Bigl( \, \pi_1 \, {\bm \mu} (s) \, \frac{1}{1-G (s; \Psi)} \, \Bigr)
\otimes \frac{1}{1-G(s; \Psi)} +O(\Delta s^2) \,.
\end{split}
\end{equation}
We have thus shown~\eqref{for-identity-1}, assuming~\eqref{assumption-1}.
This completes the proof of~\eqref{cohomomorphism-identity}.

Since we have shown~\eqref{cohomomorphism-identity} for any $s_1$ and $s_2$,
let us now take a derivative with respect to $s_2$ to obtain
\begin{equation}
{\bf H} \, {\bm \mu} (s_2) \, \frac{1}{1-\Psi}
= \frac{1}{1-H(\Psi)}
\otimes \Bigl( \, \pi_1 \, {\bf H} \, {\bm \mu} (s_2) \, \frac{1}{1-\Psi} \, \Bigr)
\otimes \frac{1}{1-H(\Psi)} \,.
\end{equation}
This holds for arbitrary ${\bm \mu} (s_2)$,
so we can replace ${\bm \mu} (s_2)$ by ${\bf C}$
and obtain~\eqref{cohomomorphism-coderivation-identity}.
Since everything other than ${\bf C}$ is degree even,
the relation holds when ${\bf C}$ is degree odd as well.
This completes the proof of~\eqref{cohomomorphism-coderivation-identity}.

Finally, let us consider~\eqref{H-field-redefinition}.
When $s_1 = 0$ and $s_2 = 1$, we have
\begin{equation}
\omega_L \Bigl( \, \pi_1 {\bf C}_1 \, \frac{1}{1 -\Psi},
\pi_1 {\bf C}_2 \, \frac{1}{1 -\Psi} \, \Bigr)
= \omega_L \Bigl( \, \pi_1 {\bf G} \, {\bf C}_1 \, \frac{1}{1 -\Psi},
\pi_1 {\bf G} \, {\bf C}_2 \, \frac{1}{1 -\Psi} \, \Bigr)
\label{G-field-redefinition}
\end{equation}
with
\begin{equation}
{\bf G} = \mathcal{P} \exp \biggl[ \, \int_0^1 ds' \, {\bm \mu} (s') \biggr] \,.
\end{equation}
We use this relation in this paper.
The relation~\eqref{H-field-redefinition} can be shown by proving
\begin{equation}
\frac{\partial}{\partial s_1} \, \omega_L \Bigl( \,
\pi_1 {\bf H} \, {\bf C}_1 \, \frac{1}{1 -\Psi},
\pi_1 {\bf H} \, {\bf C}_2 \, \frac{1}{1 -\Psi} \, \Bigr) = 0 \,.
\end{equation}
Since
\begin{equation}
\frac{\partial}{\partial s_1} \, {\bf H} = {}-{\bm \mu} (s_1) \, {\bf H}
= {}-\sum_{n=0}^\infty s_1^n \, {\bm \mu}_{n+2} \, {\bf H} \,,
\end{equation}
it is sufficient to show
\begin{equation}
\omega_L \Bigl( \,
\pi_1 \, {\bm \mu}_n \, {\bf H} \, {\bf C}_1 \, \frac{1}{1 -\Psi},
\pi_1 {\bf H} \, {\bf C}_2 \, \frac{1}{1 -\Psi} \, \Bigr)
= {}-\omega_L \Bigl( \,
\pi_1 {\bf H} \, {\bf C}_1 \, \frac{1}{1 -\Psi},
\pi_1 \, {\bm \mu}_n \, {\bf H} \, {\bf C}_2 \, \frac{1}{1 -\Psi} \, \Bigr)
\label{move-mu_n}
\end{equation}
for any $n$.
The left-hand side of~\eqref{move-mu_n} is
\begin{align}
& \omega_L \Bigl( \,
\pi_1 \, {\bm \mu}_n \, {\bf H} \, {\bf C}_1 \, \frac{1}{1 -\Psi},
\pi_1 {\bf H} \, {\bf C}_2 \, \frac{1}{1 -\Psi} \, \Bigr)
= \omega_L \Bigl( \,
\pi_1 \, {\bm \mu}_n \, \frac{1}{1-H(\Psi)} \, H_{{\bf C}_1} (\Psi) \, \frac{1}{1-H(\Psi)},
H_{{\bf C}_2} (\Psi) \, \Bigr) \nonumber \\
& = \sum_{k=0}^{n-1} \omega_L \Bigl( \,
\mu_n ( \, \underbrace{H(\Psi), \ldots , H(\Psi)}_k \,, H_{{\bf C}_1} (\Psi),
\underbrace{H(\Psi), \ldots , H(\Psi)}_{n-1-k} \, ),
H_{{\bf C}_2} (\Psi) \, \Bigr) \,,
\end{align}
where
\begin{equation}
H_{\bf C} (\Psi) = \pi_1 \, {\bf H} \, {\bf C} \, \frac{1}{1-\Psi} \,.
\end{equation}
Note that $H(\Psi)$ is degree even.
When $A$ is degree even, we have
\begin{equation}
\omega_L \, ( \, \mu_n ( A, A_1, \ldots , A_{n-1} ), A_n \, )
= {}-\omega_L \, ( A, \mu_n ( A_1, \ldots , A_n ) \, )
= \omega_L \, ( \, \mu_n ( A_1, \ldots , A_n ), A \, ) \,.
\end{equation}
We use this formula to find
\begin{equation}
\begin{split}
& \sum_{k=0}^{n-1} \omega_L \Bigl( \,
\mu_n ( \, \underbrace{H(\Psi), \ldots , H(\Psi)}_k \,, H_{{\bf C}_1} (\Psi),
\underbrace{H(\Psi), \ldots , H(\Psi)}_{n-1-k} \, ),
H_{{\bf C}_2} (\Psi) \, \Bigr) \\
& = \omega_L \Bigl( \,
\mu_n ( \, H_{{\bf C}_1} (\Psi),
\underbrace{H(\Psi), \ldots , H(\Psi)}_{n-1} \, ),
H_{{\bf C}_2} (\Psi) \, \Bigr) \\
& \quad~ +\sum_{k=1}^{n-1} \omega_L \Bigl( \,
\mu_n ( H_{{\bf C}_1} (\Psi),
\, \underbrace{H(\Psi), \ldots , H(\Psi)}_{n-1-k} \,, H_{{\bf C}_2} (\Psi),
\underbrace{H(\Psi), \ldots , H(\Psi)}_{k-1} \, ), H(\Psi) \, ) \, \Bigr) \\
& = -\sum_{k=0}^{n-1} \omega_L \Bigl( \, H_{{\bf C}_1} (\Psi),
\mu_n ( \, \underbrace{H(\Psi), \ldots , H(\Psi)}_{n-1-k} \,, H_{{\bf C}_2} (\Psi),
\underbrace{H(\Psi), \ldots , H(\Psi)}_k \, ) \, \Bigr) \\
& = {}-\omega_L \Bigl( \,
\pi_1 {\bf H} \, {\bf C}_1 \, \frac{1}{1 -\Psi},
\pi_1 \, {\bm \mu}_n \, {\bf H} \, {\bf C}_2 \, \frac{1}{1 -\Psi} \, \Bigr) \,.
\end{split}
\end{equation}
We have thus shown the relation~\eqref{move-mu_n}.
This completes the proof of~\eqref{H-field-redefinition}.

\section{Equivalence of the actions up to quartic interactions}
\setcounter{equation}{0}
\label{up-to-quartic}

In this appendix we demonstrate the equivalence of
the theory with the $A_\infty$ structure constructed in~\cite{Erler:2013xta}
and the theory obtained from the Berkovits formulation
by partial gauge fixing~\cite{Iimori:2013kha}
up to quartic interactions by constructing the field redefinition explicitly.
We also translate the descriptions of the two-string products and the three-string products
into the standard language in string field theory.

\subsection{The expansion of the multi-string products}

We have seen in subsection~\ref{A_infinity-theory} that the coderivation ${\bf M} (s)$
can be written as
\begin{equation}
{\bf M} (s) = {\bf G}^{-1} (s) \, {\bf Q} \, {\bf G} (s) 
\label{M(s)}
\end{equation}
with ${\bf G} (s)$ satisfying
\begin{equation}
[ \, {\bm \eta}, {\bf G} (s) \, ]
= s \, {\bf m}_2 \, {\bf G} (s) \,.
\label{eta-G(s)}
\end{equation}
In this subsection, we present an explicit expression of ${\bf M} (s)$ expanded in $s$ to $O(s^3)$
and demonstrate that these relations are satisfied.

Let us first solve the differential equation
\begin{equation}
\frac{d}{ds} \, {\bf M} (s) = [ \, {\bf M} (s), {\bm \mu} (s) \, ]
\label{M(s)-equation}
\end{equation}
with the initial condition
\begin{equation}
{\bf M} (0) = {\bf Q}
\end{equation}
by expanding ${\bf M} (s)$ and ${\bm \mu} (s)$ in $s$ as follows:
\begin{align}
{\bf M} (s) & = \sum_{n=0}^{\infty} s^n \, {\bf M}_{n+1}
= {\bf Q} + s \, {\bf M}_2 + s^2 \, {\bf M}_3 + s^3 \, {\bf M}_4 +O(s^4) \,, \\
{\bm \mu} (s) & = \sum_{n=0}^{\infty} s^n \, {\bm \mu}_{n+2}
= {\bm \mu}_2 + s \, {\bm \mu}_3 + s^2 \, {\bm \mu}_4 +O(s^3) \,.
\label{mu-expansion}
\end{align}
Since
\begin{equation}
\frac{d}{ds} \, {\bf M} (s) = {\bf M}_2 + 2 \, s \, {\bf M}_3 + 3 \, s^2 \, {\bf M}_4 +O(s^3) \,,
\end{equation}
the differential equation~\eqref{M(s)-equation} expanded in $s$ to $O(s^2)$ gives
\begin{equation}
\begin{split}
{\bf M}_2 & = [ \, {\bf Q}, {\bm\mu}_2 \, ] \,, \\
{\bf M}_3 & = \frac{1}{2} \, [ \, {\bf Q}, {\bm \mu}_3 \, ]
+\frac{1}{2} \, [ \, {\bf M}_2, {\bm \mu}_2 \, ] \,, \\
{\bf M}_4 & = \frac{1}{3} \, [ \, {\bf Q}, {\bm \mu}_4 \, ] 
+\frac{1}{3} \, [ \, {\bf M}_2, {\bm \mu}_3 \, ]
+\frac{1}{3} \, [ \, {\bf M}_3 , {\bm \mu}_2 \, ] \,.
\end{split}
\label{explicit-M_2}
\end{equation}
As can be seen from these equations, the differential equation~\eqref{M(s)-equation}
determines the form of ${\bf M}_n$ written in terms of ${\bf Q}$, ${\bm \mu}_2$,
${\bm \mu}_3$, \ldots \,, ${\bm \mu}_n$.
For ${\bf M}_3$ and ${\bf M}_4$,  we find
\begin{equation}
\begin{split}
{\bf M}_3 & = \frac{1}{2} \, [ \, {\bf Q}, {\bm \mu}_3 \, ]
+ \frac{1}{2} \, [ \, [{\bf Q}, {\bm\mu}_2 \, ], {\bm \mu}_2 \, ] \,, \\
{\bf M}_4 & = \frac{1}{3} \, [ \, {\bf Q}, {\bm \mu}_4 \, ] 
+\frac{1}{3} \, [ \, [ \, {\bf Q}, {\bm\mu}_2 \, ], {\bm \mu}_3 \, ]
+\frac{1}{6} \, [ \, [ \, {\bf Q}, {\bm \mu}_3 \, ], {\bm \mu}_2 \, ]  
+\frac{1}{6} \, [ \, [ \, [ \, {\bf Q}, {\bm\mu}_2 \, ], {\bm \mu}_2 \, ], {\bm \mu}_2 \, ] \,.
\end{split}
\label{explicit-M_3-M_4}
\end{equation}

Let us compare these expressions for ${\bf M}_2$, ${\bf M}_3$, and ${\bf M}_4$
with the form~\eqref{M(s)}.
The path-ordered exponential ${\bf G} (s)$ can be expanded in $s$ as follows:
\begin{equation}
\begin{split}
{\bf G} (s) & = \mathcal{P} \exp \biggl[ \, \int_0^s ds' \, {\bm \mu} (s') \, \biggr] \\
& = 1 +\int_0^s ds_1 \, {\bm \mu} (s_1) 
+\int_0^s ds_1 \int_0^{s_1} ds_2 \, {\bm \mu} (s_2) \, {\bm \mu} (s_1) \\
& \quad~ + \int_0^s ds_1 \int_0^{s_1} ds_2 \int_0^{s_2} ds_3 \,
{\bm \mu} (s_3) \, {\bm \mu} (s_2) \, {\bm \mu} (s_1) +O(s^4) \\
& = 1 +s \, {\bm \mu}_2  
+s^2 \, \biggl( \, \frac{1}{2} \, {\bm \mu}_3
+\frac{1}{2} \, {\bm \mu}_2 \, {\bm \mu}_2 \biggr) \\
& \quad~ +s^3 \, \biggl( \, \frac{1}{3} \, {\bm \mu}_4
+\frac{1}{3} \, {\bm \mu}_2 \, {\bm \mu}_3 
+\frac{1}{6} \, {\bm \mu}_3 \, {\bm \mu}_2 
+\frac{1}{6} \,  {\bm \mu}_2 \, {\bm \mu}_2 \, {\bm \mu}_2 \biggr) +O(s^4) \,.
\end{split}
\label{G(s)-expansion}
\end{equation}
Its inverse ${\bf G}^{-1} (s)$ is
\begin{equation}
\begin{split}
{\bf G}^{-1} (s) & = \mathcal{P} \exp \biggl[ \, \int_s^0 ds' \, {\bm \mu} (s') \, \biggr] \\
& = 1 +\int_s^0 ds_1 \, {\bm \mu} (s_1) 
+\int_s^0 ds_1 \int_{s_1}^0 ds_2 \, {\bm \mu} (s_1) \, {\bm \mu} (s_2) \\
& \quad~ + \int_s^0 ds_1 \int_{s_1}^0 ds_2 \int_{s_2}^0 ds_3 \,
{\bm \mu} (s_1) \, {\bm \mu} (s_2) \, {\bm \mu} (s_3) +O(s^4) \\
& = 1 -s \, {\bm \mu}_2  
+s^2 \, \biggl( \, -\frac{1}{2} \, {\bm \mu}_3
+\frac{1}{2} \, {\bm \mu}_2 \, {\bm \mu}_2 \biggr) \\
& \quad~ +s^3 \, \biggl( \, -\frac{1}{3} \, {\bm \mu}_4
+\frac{1}{6} \, {\bm \mu}_2 \, {\bm \mu}_3 
+\frac{1}{3} \, {\bm \mu}_3 \, {\bm \mu}_2 
-\frac{1}{6} \,  {\bm \mu}_2 \, {\bm \mu}_2 \, {\bm \mu}_2 \biggr) +O(s^4) \,.
\end{split}
\end{equation}
Then the expansion of ${\bf G}^{-1} (s) \, {\bf Q} \, {\bf G} (s)$ is given by
\begin{equation}
{\bf G}^{-1} (s) \, {\bf Q} \, {\bf G} (s)
= {\bf Q} +s \, {\bf N}_2 +s^2 \, {\bf N}_3 +s^3 \, {\bf N}_4 +O(s^4) \,,
\end{equation}
where
\begin{equation}
\begin{split}
{\bf N}_2 & = {\bf Q} \, {\bm \mu}_2 -{\bm \mu}_2 \, {\bf Q}
= [ \, {\bf Q}, {\bm \mu}_2 \, ] \,, \\
{\bf N}_3 & = {\bf Q} \, \biggl( \, \frac{1}{2} \, {\bm \mu}_3
+\frac{1}{2} \, {\bm \mu}_2 \, {\bm \mu}_2 \biggr)
-{\bm \mu}_2 \, {\bf Q} \, {\bm \mu}_2
+\biggl( \, -\frac{1}{2} \, {\bm \mu}_3
+\frac{1}{2} \, {\bm \mu}_2 \, {\bm \mu}_2 \biggr) \, {\bf Q} \\
& = \frac{1}{2} \, [ \, {\bf Q}, {\bm \mu}_3 \, ]
+ \frac{1}{2} \, [ \, [ \, {\bf Q}, {\bm\mu}_2 \, ], {\bm \mu}_2 \, ] \,, \\
{\bf N}_4 & = {\bf Q} \, \biggl( \, \frac{1}{3} \, {\bm \mu}_4
+\frac{1}{3} \, {\bm \mu}_2 \, {\bm \mu}_3 
+\frac{1}{6} \, {\bm \mu}_3 \, {\bm \mu}_2 
+\frac{1}{6} \,  {\bm \mu}_2 \, {\bm \mu}_2 \, {\bm \mu}_2 \biggr)
-{\bm \mu}_2 \, {\bf Q} \, \biggl( \, \frac{1}{2} \, {\bm \mu}_3
+\frac{1}{2} \, {\bm \mu}_2 \, {\bm \mu}_2 \biggr) \\
& \quad~ +\biggl( \, -\frac{1}{2} \, {\bm \mu}_3
+\frac{1}{2} \, {\bm \mu}_2 \, {\bm \mu}_2 \biggr) \, {\bf Q} \, {\bm \mu}_2
+\biggl( \, -\frac{1}{3} \, {\bm \mu}_4
+\frac{1}{6} \, {\bm \mu}_2 \, {\bm \mu}_3 
+\frac{1}{3} \, {\bm \mu}_3 \, {\bm \mu}_2 
-\frac{1}{6} \,  {\bm \mu}_2 \, {\bm \mu}_2 \, {\bm \mu}_2 \biggr) \, {\bf Q} \\
& = \frac{1}{3} \, [ \, {\bf Q}, {\bm \mu}_4 \, ] 
+\frac{1}{3} \, [ \, [ \, {\bf Q}, {\bm\mu}_2 \, ], {\bm \mu}_3 \, ]
+\frac{1}{6} \, [ \, [ \, {\bf Q}, {\bm \mu}_3 \, ], {\bm \mu}_2 \, ]  
+\frac{1}{6} \, [ \, [ \, [ \, {\bf Q}, {\bm\mu}_2 \, ], {\bm \mu}_2 \, ], {\bm \mu}_2 \, ] \,.
\end{split}
\end{equation}
We have thus verified that ${\bf M}_2$, ${\bf M}_3$, and ${\bf M}_4$
in~\eqref{explicit-M_2} and~\eqref{explicit-M_3-M_4}
are reproduced by the expansion of ${\bf G}^{-1} (s) \, {\bf Q} \, {\bf G} (s)$.

Let us next consider the coderivation ${\bm \mu} (s)$.
The conditions on ${\bm \mu} (s)$ are characterized by
\begin{equation}
[ \, {\bm \eta}, {\bm \mu} (s) \, ] = {\bf m} (s)
\label{eta-mu}
\end{equation}
and the differential equation for ${\bf m} (s)$
\begin{equation}
\frac{d}{ds} \, {\bf m}(s) = [ \, {\bf m}(s), {\bm \mu}(s) \, ]
\label{m(s)-equation}
\end{equation}
with the initial condition
\begin{equation}
{\bf m} (0) = {\bf m}_2 \,.
\end{equation}
We expand ${\bm \mu} (s)$ as in~\eqref{mu-expansion}
and ${\bf m} (s)$ as
\begin{equation}
{\bf m} (s) = \sum_{n=0}^{\infty} s^n \, {\bf m}_{n+2}
= {\bf m}_2 +s \, {\bf m}_3  +s^2 \, {\bf m}_4 +O(s^3)
\end{equation}
to obtain
\begin{align}
[ \, {\bm \eta}, {\bm \mu}_2 \, ] & = {\bf m}_2 \,,
\label{mu_2-condition} \\
{\bf m}_3 & = [ \, {\bf m}_2, {\bm\mu}_2 \, ] \,,
\label{determine-m_3} \\
[ \, {\bm \eta}, {\bm \mu}_3 \, ] & = {\bf m}_3 \,,
\label{mu_3-condition} \\
{\bf m}_4 & = \frac{1}{2} \, [ \, {\bf m}_2, {\bm \mu}_3 \, ]
+\frac{1}{2} \, [ \, {\bf m}_3, {\bm \mu}_2] \,,
\label{determine-m_4} \\
[ \, {\bm \eta}, {\bm \mu}_4 \, ] & = {\bf m}_4 \,.
\label{mu_4-condition}
\end{align}
The coderivation ${\bm \mu}_2$ is constrained to satisfy~\eqref{mu_2-condition}.
The cohomology of $\eta$ is trivial,
and we can use $\xi$ to construct ${\bm \mu}_2$ satisfying~\eqref{mu_2-condition}.
We present an explicit realization of ${\bm \mu}_2$ in the next subsection.
Once we decide on ${\bm \mu}_2$, the coderivation ${\bf m}_3$ is determined
by~\eqref{determine-m_3}.
Then the coderivation ${\bm \mu}_3$ is constrained to satisfy~\eqref{mu_3-condition}.
The cohomology of $\eta$ is again trivial,
and we can use $\xi$ to construct ${\bm \mu}_3$ satisfying~\eqref{mu_3-condition}.
We present an explicit realization of ${\bm \mu}_3$ in the next subsection.
This way we can construct ${\bm \mu}_n$ and ${\bf m}_n$ from~\eqref{eta-mu}
and~\eqref{m(s)-equation}.

Finally, let us verify that the relation~\eqref{eta-G(s)} is satisfied
for the expansion of ${\bf G} (s)$ given in~\eqref{G(s)-expansion}.
Using~\eqref{mu_2-condition}, \eqref{mu_3-condition}, and~\eqref{mu_4-condition}, 
the commutator $[ \, {\bm \eta}, {\bf G} (s) \, ]$ is given by
\begin{equation}
\begin{split}
[ \, {\bm \eta}, {\bf G} (s) \, ] 
& = s \, {\bf m}_2  
+s^2 \, \biggl( \, \frac{1}{2} \, {\bf m}_3
+\frac{1}{2} \, ( \, {\bf m}_2 \, {\bm \mu}_2
+{\bm \mu}_2 \, {\bf m}_2 \, ) \, \biggr) \\
& \quad~ +s^3 \, \biggl( \, \frac{1}{3} \, {\bf m}_4
+\frac{1}{3} \, ( \, {\bf m}_2 \, {\bm \mu}_3
+{\bm \mu}_2 \, {\bf m}_3 \, )
+\frac{1}{6} \, ( \, {\bf m}_3 \, {\bm \mu}_2
+{\bm \mu}_3 \, {\bf m}_2 \, ) \\
& \qquad \qquad \quad 
+\frac{1}{6} \,  ( \, {\bf m}_2 \, {\bm \mu}_2 \, {\bm \mu}_2
+{\bm \mu}_2 \, {\bf m}_2 \, {\bm \mu}_2
+{\bm \mu}_2 \, {\bm \mu}_2 \, {\bf m}_2 \, ) \, \biggr) +O(s^4) \,.
\end{split}
\end{equation}
Using the relations~\eqref{determine-m_3} and~\eqref{determine-m_4}, we find
\begin{align}
& \frac{1}{2} \, {\bf m}_3
+\frac{1}{2} \, ( \, {\bf m}_2 \, {\bm \mu}_2
+{\bm \mu}_2 \, {\bf m}_2 \, )
= \frac{1}{2} \, ( \, {\bf m}_2 \, {\bm \mu}_2
-{\bm \mu}_2 \, {\bf m}_2 \, )
+\frac{1}{2} \, ( \, {\bf m}_2 \, {\bm \mu}_2
+{\bm \mu}_2 \, {\bf m}_2 \, )
= {\bf m}_2 \, {\bm \mu}_2 \,, \\
& \frac{1}{3} \, {\bf m}_4
+\frac{1}{3} \, ( \, {\bf m}_2 \, {\bm \mu}_3
+{\bm \mu}_2 \, {\bf m}_3 \, )
+\frac{1}{6} \, ( \, {\bf m}_3 \, {\bm \mu}_2
+{\bm \mu}_3 \, {\bf m}_2 \, )
+\frac{1}{6} \,  ( \, {\bf m}_2 \, {\bm \mu}_2 \, {\bm \mu}_2
+{\bm \mu}_2 \, {\bf m}_2 \, {\bm \mu}_2
+{\bm \mu}_2 \, {\bm \mu}_2 \, {\bf m}_2 \, ) \nonumber \\
& = {\bf m}_2 \, \biggl( \, \frac{1}{2} \, {\bm \mu}_3
+\frac{1}{2} \, {\bm \mu}_2 \, {\bm \mu}_2 \biggr) \,.
\end{align}
We have thus confirmed~\eqref{eta-G(s)} up to $O(s^4)$.

\subsection{Translation into string field theory conventions}

While the grading with respect to degree is convenient and efficient
for describing the $A_\infty$ structure,
it is relatively unfamiliar
and it would be useful to translate the results of this paper
into standard conventions used in string field theory.
We consider two-string products and three-string products in the rest of this appendix.

In subsection~\ref{A_infinity-review},
we have related $M_2$, $M_3$, and $m_2$
to $V_2$, $V_3$, and the star product, respectively.
For $m_3$, $\mu_2$, and $\mu_3$,
we introduce $V_3^\xi$, $V_2^\xi$, and $V_3^{\xi \xi}$, respectively.
The relations are summarized as follows:
\begin{equation}
\begin{split}
M_2 (A_1, A_2) & = (-1)^{{\rm deg} (A_1)} V_2 (A_1, A_2) \,, \\
M_3 (A_1, A_2, A_3) & = (-1)^{{\rm deg} (A_2)} V_3 (A_1, A_2, A_3) \,, \\
m_2 (A_1, A_2) & = (-1)^{{\rm deg} (A_1)} A_1 \, A_2 \,, \\
m_3 (A_1, A_2, A_3) & = (-1)^{{\rm deg} (A_2)} V_3^\xi (A_1, A_2, A_3) \,, \\
\mu_2 (A_1, A_2) & = (-1)^{{\rm deg} (A_1)} V_2^\xi (A_1, A_2) \,, \\
\mu_3 (A_1, A_2, A_3) & = (-1)^{{\rm deg} (A_2)} V_3^{\xi \xi} (A_1, A_2, A_3) \,.
\end{split}
\end{equation}
The cyclic properties are translated into
\begin{equation}
\begin{split}
\langle \, A_1, V_2 \, (A_2, A_3) \, \rangle
& = \langle \, V_2 \, (A_1, A_2), A_3 \, \rangle \,, \\
\langle \, A_1, V_3 \, (A_2, A_3, A_4) \, \rangle
& = {}-(-1)^{A_1} \langle \, V_3 \, (A_1, A_2, A_3), A_4 \, \rangle \,, \\
\langle \, A_1, A_2 \, A_3 \, \rangle
& = \langle \, A_1 \, A_2, A_3 \, \rangle \,, \\
\langle \, A_1, V_3^\xi \, (A_2, A_3, A_4) \, \rangle
& = {}-(-1)^{A_1} \langle \, V_3^\xi \, (A_1, A_2, A_3), A_4 \, \rangle \,, \\
\langle \, A_1, V_2^\xi \, (A_2, A_3) \, \rangle
& = (-1)^{A_1} \langle \, V_2^\xi \, (A_1, A_2), A_3 \, \rangle \,, \\
\langle \, A_1, V_3^{\xi \xi} \, (A_2, A_3, A_4) \, \rangle
& = {}-\langle \, V_3^{\xi \xi} \, (A_1, A_2, A_3), A_4 \, \rangle \,.
\end{split}
\end{equation}

The condition~\eqref{mu_2-condition}
\begin{equation}
\eta \mu_2 (A_1, A_2) -\mu_2 (\eta A_1, A_2) -(-1)^{{\rm deg} (A_1)} \mu_2 (A_1, \eta A_2)
= m_2 (A_1, A_2)
\end{equation}
on $\mu_2$ is translated into the following condition on $V_2^\xi$:
\begin{equation}
\eta V_2^\xi (A_1, A_2) +V_2^\xi (\eta A_1, A_2) +(-1)^{A_1} V_2^\xi (A_1, \eta A_2)
= A_1 A_2 \,.
\end{equation}
One realization of $V_2^\xi (A_1, A_2)$ is
\begin{equation}
V_2^\xi (A_1, A_2) = \frac{1}{3} \, \Bigl[ \,
\xi ( A_1 A_2 ) +(\xi A_1) A_2 +(-1)^{A_1} A_1 (\xi A_2) \, \Bigr] \,.
\label{V_2^xi}
\end{equation}
The expression of ${\bf M}_2$ in~\eqref{explicit-M_2}
is translated into the following expression of the two-string product $V_2 (A_1, A_2)$:
\begin{equation}
V_2 (A_1, A_2)
= Q V_2^\xi (A_1, A_2) +V_2^\xi (Q A_1, A_2) +(-1)^{A_1} V_2^\xi (A_1, Q A_2) \,.
\end{equation}
With the choice of $V_2^\xi (A_1, A_2)$ in~\eqref{V_2^xi},
$V_2 (A_1, A_2)$ is
\begin{equation}
V_2 (A_1, A_2)
= \frac{1}{3} \, \Bigl[ \, X ( A_1 A_2 ) +(X A_1) A_2 +A_1 (X A_2) \, \Bigr] \,,
\end{equation}
where
\begin{equation}
X = \{ \, Q, \xi \, \} \,.
\end{equation}
When we use a line integral of $\xi (z)$ to realize $\xi$,
the operator $X$ takes the form of a line integral of the picture-changing operator.
The cubic interaction written in terms of $V_2$ with this choice
does not suffer from singularities coming from local insertions of the picture-changing operator.

The three-string product $m_3$ is determined by~\eqref{determine-m_3}:
\begin{equation}
\begin{split}
m_3 (A_1, A_2, A_3)
& = m_2 ( \mu_2 (A_1, A_2), A_3 ) +m_2 ( A_1, \mu_2 (A_2, A_3) ) \\
& \quad~ -\mu_2 ( m_2 (A_1, A_2), A_3) -(-1)^{{\rm deg} (A_1)} \mu_2 ( A_1, m_2 (A_2, A_3) ) \,.
\end{split}
\end{equation}
This is translated into the following expression for $V_3^\xi$ in terms of $V_2^\xi$:
\begin{equation}
V_3^\xi (A_1, A_2, A_3)
= V_2^\xi (A_1, A_2) \, A_3 -(-1)^{A_1} A_1 \, V_2^\xi (A_2, A_3)
+V_2^\xi (A_1 A_2, A_3) -V_2^\xi (A_1, A_2 A_3) \,.
\end{equation}
With the choice of $V_2^\xi$ in~\eqref{V_2^xi},
$V_3^\xi (A_1, A_2, A_3)$ is
\begin{equation}
V_3^\xi (A_1, A_2, A_3) = \frac{2}{3} \, \Bigl[ \,
( \xi (A_1 A_2) ) \, A_3 -(-1)^{A_1} A_1 \, ( \xi (A_2 A_3) ) \, \Bigr] \,.
\end{equation}

The condition~\eqref{mu_3-condition}
\begin{equation}
\begin{split}
& \eta \mu_3 (A_1, A_2, A_3)
-\mu_3 (\eta A_1, A_2, A_3)
-(-1)^{{\rm deg} (A_1)} \mu_3 (A_1, \eta A_2, A_3) \\
& -(-1)^{{\rm deg} (A_1)+{\rm deg} (A_2)} \mu_3 (A_1, A_2, \eta A_3)
= m_3 (A_1, A_2, A_3)
\end{split}
\end{equation}
on $\mu_3$ is translated into the following condition on $V_3^{\xi \xi}$:
\begin{align}
& \eta V_3^{\xi \xi} (A_1, A_2, A_3) 
-V_3^{\xi \xi} (\eta A_1, A_2, A_3)
-(-1)^{A_1} V_3^{\xi \xi} (A_1, \eta A_2, A_3)
-(-1)^{A_1+A_2} V_3^{\xi \xi} (A_1, A_2, \eta A_3) \nonumber \\
& = V_3^\xi (A_1, A_2, A_3) \,.
\end{align}
Since $V_3^\xi$ is written in terms of $V_2^\xi$,
this condition can be understood as the following condition
between $V_3^{\xi \xi}$ and $V_2^\xi$:
\begin{align}
& \eta V_3^{\xi \xi} (A_1, A_2, A_3) 
-V_3^{\xi \xi} (\eta A_1, A_2, A_3)
-(-1)^{A_1} V_3^{\xi \xi} (A_1, \eta A_2, A_3)
-(-1)^{A_1+A_2} V_3^{\xi \xi} (A_1, A_2, \eta A_3) \nonumber \\
& = V_2^\xi (A_1, A_2) \, A_3 -(-1)^{A_1} A_1 \, V_2^\xi (A_2, A_3)
+V_2^\xi (A_1 A_2, A_3) -V_2^\xi (A_1, A_2 A_3) \,.
\end{align}
When $V_3^\xi (A_1, A_2, A_3)$ is given,
one realization of $V_3^{\xi \xi} (A_1, A_2, A_3)$ is
\begin{equation}
\begin{split}
V_3^{\xi \xi} (A_1, A_2, A_3) & = \frac{1}{4} \, \Bigl[ \,
\xi V_3^\xi (A_1, A_2, A_3)
-V_3^\xi (\xi A_1, A_2, A_3) \\
& \qquad \quad -(-1)^{A_1} V_3^\xi (A_1, \xi A_2, A_3)
-(-1)^{A_1+A_2} V_3^\xi (A_1, A_2, \xi A_3) \, \Bigr] \,.
\label{V_3^xi-xi}
\end{split}
\end{equation}
With the choices of $V_2^\xi$ in~\eqref{V_2^xi}
and $V_3^{\xi \xi}$ in~\eqref{V_3^xi-xi},
the explicit expression for $V_3^{\xi \xi} (A_1, A_2, A_3)$ is given by
\begin{equation}
\begin{split}
V_3^{\xi \xi} (A_1, A_2, A_3) & = \frac{1}{6} \, \Bigl[ \,
\xi ( ( \xi (A_1 A_2) ) \, A_3 ) -(-1)^{A_1} \xi ( A_1 \, ( \xi (A_2 A_3) ) ) \\
& \qquad \quad
-( \xi ( (\xi A_1) A_2) ) ) \, A_3 -(-1)^{A_1} (\xi A_1) \, ( \xi (A_2 A_3) ) \\
& \qquad \quad
-(-1)^{A_1} ( \xi (A_1 (\xi A_2) ) ) \, A_3 +A_1 \, ( \xi ( (\xi A_2) A_3) ) \\
& \qquad \quad
-(-1)^{A_1+A_2} ( \xi (A_1 A_2) ) \, (\xi A_3) +(-1)^{A_2} A_1 \, ( \xi (A_2 (\xi A_3) ) ) \, \Bigr] \,.
\end{split}
\end{equation}
The expression of ${\bf M}_3$ in~\eqref{explicit-M_2}
is translated into the following expression of the three-string product $V_3 (A_1, A_2, A_3)$:
\begin{equation}
\begin{split}
V_3 (A_1, A_2, A_3)
& = \frac{1}{2} \, \Bigl[ \,
Q V_3^{\xi \xi} (A_1, A_2, A_3)
-V_3^{\xi \xi} (Q A_1, A_2, A_3) \\
& \qquad \quad -(-1)^{A_1} V_3^{\xi \xi} (A_1, Q A_2, A_3)
-(-1)^{A_1+A_2} V_3^{\xi \xi} (A_1, A_2, Q A_3) \\
& \qquad \quad +V_2 (V_2^\xi (A_1, A_2), A_3)
-(-1)^{A_1} V_2 (A_1, V_2^\xi (A_2, A_3)) \\
& \qquad \quad +V_2^\xi (V_2 (A_1, A_2), A_3)
-V_2^\xi (A_1, V_2 (A_2, A_3)) \, \Bigr] \,. \\
\end{split}
\end{equation}
Since $V_2$ is written in terms of $V_2^\xi$,
the three-string product $V_3$ is given
in terms of $V_2^\xi$ and $V_3^{\xi \xi}$ by
\begin{equation}
\begin{split}
V_3 (A_1, A_2, A_3)
& = \frac{1}{2} \, \Bigl[ \,
Q V_3^{\xi \xi} (A_1, A_2, A_3)
-V_3^{\xi \xi} (Q A_1, A_2, A_3) \\
& \qquad \quad -(-1)^{A_1} V_3^{\xi \xi} (A_1, Q A_2, A_3)
-(-1)^{A_1+A_2} V_3^{\xi \xi} (A_1, A_2, Q A_3) \\
& \qquad \quad
+Q V_2^\xi (V_2^\xi (A_1, A_2), A_3)
-(-1)^{A_1} Q V_2^\xi (A_1, V_2^\xi (A_2, A_3)) \\
& \qquad \quad
+2 \, V_2^\xi (Q V_2^\xi (A_1, A_2), A_3)
-2 \, V_2^\xi (A_1, Q V_2^\xi (A_2, A_3)) \\
& \qquad \quad
+V_2^\xi (V_2^\xi (Q A_1, A_2), A_3)
+(-1)^{A_1} V_2^\xi (V_2^\xi (A_1, Q A_2), A_3) \\
& \qquad \quad
-(-1)^{A_1+A_2} V_2^\xi (V_2^\xi (A_1, A_2), Q A_3)
-(-1)^{A_1} V_2^\xi (Q A_1, V_2^\xi (A_2, A_3)) \\
& \qquad \quad
-V_2^\xi (A_1, V_2^\xi (Q A_2, A_3))
-(-1)^{A_2} V_2^\xi (A_1, V_2^\xi (A_2, Q A_3)) \,
\Bigr] \,. \\
\end{split}
\end{equation}
This corresponds to the expression of ${\bf M}_3$ in~\eqref{explicit-M_3-M_4}.
For a Grassmann-odd string field $\Psi$,
the three-string product $V_3 (\Psi, \Psi, \Psi)$ is
\begin{equation}
\begin{split}
V_3 (\Psi, \Psi, \Psi)
& = \frac{1}{2} \, \Bigl[ \,
Q V_3^{\xi \xi} (\Psi, \Psi, \Psi)
-V_3^{\xi \xi} (Q \Psi, \Psi, \Psi)
+V_3^{\xi \xi} (\Psi, Q \Psi, \Psi)
-V_3^{\xi \xi} (\Psi, \Psi, Q \Psi) \\
& \qquad \quad
+Q V_2^\xi (V_2^\xi (\Psi, \Psi), \Psi)
+Q V_2^\xi (\Psi, V_2^\xi (\Psi, \Psi)) \\
& \qquad \quad
+2 \, V_2^\xi (Q V_2^\xi (\Psi, \Psi), \Psi)
-2 \, V_2^\xi (\Psi, Q V_2^\xi (\Psi, \Psi)) \\
& \qquad \quad
+V_2^\xi (Q \Psi, V_2^\xi (\Psi, \Psi))
-V_2^\xi (V_2^\xi (\Psi, Q \Psi), \Psi)
-V_2^\xi (V_2^\xi (\Psi, \Psi), Q \Psi) \\
& \qquad \quad
+V_2^\xi (Q V_2^\xi (\Psi, \Psi), \Psi)
-V_2^\xi (\Psi, V_2^\xi (Q \Psi, \Psi))
+V_2^\xi (\Psi, V_2^\xi (\Psi, Q \Psi)) \, \Bigr] \,.
\end{split}
\label{V_3(Psi,Psi,Psi)}
\end{equation}

\subsection{The action up to quartic interactions}

In the rest of this appendix, we rescale $\Psi$ to $g \Psi$,
where $g$ is the coupling constant, and multiply the action by $1/g^2$.
The action of open superstring field theory with the $A_\infty$ structure
can be written as
\begin{equation}
S = S_2 +g \, S_3 +g^2 S_4 +O(g^3) \,,
\end{equation}
where
\begin{equation}
\begin{split}
S_2 & = {}-\frac{1}{2} \, \langle \, \xi \Psi, Q \Psi \, \rangle
= -\int_0^1 dt \, \langle \, \xi \Psi, Q t \Psi \, \rangle \,, \\
S_3 & = {}-\frac{1}{3} \, \langle \, \xi \Psi, V_2 (\Psi, \Psi) \, \rangle
= -\int_0^1 dt \, \langle \, \xi \Psi, V_2 (t \Psi, t \Psi) \, \rangle \,, \\
S_4 & = {}-\frac{1}{4} \, \langle \, \xi \Psi, V_3 (\Psi, \Psi, \Psi) \, \rangle
= -\int_0^1 dt \, \langle \, \xi \Psi, V_3 (t \Psi, t \Psi, t \Psi) \, \rangle \,.
\end{split}
\end{equation}
Following subsection~\ref{WZW-like-form}, let us transform this action
to the WZW-like form. The cubic interaction $S_3$ is
\begin{equation}
S_3 = -\int_0^1 dt \, \Bigl[ \,
\langle \, \xi \Psi, Q V_2^\xi (t \Psi, t \Psi) \, \rangle
+\langle \, \xi \Psi, V_2^\xi (Q t \Psi, t \Psi) \, \rangle
-\langle \, \xi \Psi, V_2^\xi (t \Psi, Q t \Psi) \, \rangle \, \Bigr] \,.
\end{equation}
Using the cyclic property of $V_2^\xi$, this can be written as
\begin{equation}
S_3 = -\int_0^1 dt \,
\langle \, \xi \Psi, Q V_2^\xi (t \Psi, t \Psi) \, \rangle
+\int_0^1 dt \, 
\langle \, V_2^\xi (\xi \Psi, t \Psi) -V_2^\xi (t \Psi, \xi \Psi) , Q t \Psi \, \rangle \,.
\end{equation}
The quartic interaction $S_4$ is written in terms of $V_3$ in~\eqref{V_3(Psi,Psi,Psi)}.
Using the cyclic properties of $V_2^\xi$ and $V_3^{\xi \xi}$,
$S_4$ can be written as
\begin{equation}
\begin{split}
S_4 & = {}-\frac{1}{2} \, \int_0^1 dt \, 
\langle \, \xi \Psi, Q V_3^{\xi \xi} (t \Psi, t \Psi, t \Psi)
+Q V_2^\xi (V_2^\xi (t \Psi, t \Psi), t \Psi)
+Q V_2^\xi (t \Psi, V_2^\xi (t \Psi, t \Psi)) \, \rangle \\
& \quad~ +\int_0^1 dt \, 
\langle \, V_2^\xi (\xi \Psi, t \Psi)-V_2^\xi (t \Psi, \xi \Psi), Q V_2^\xi (t \Psi, t \Psi)  \, \rangle \\
& \quad~ -\frac{1}{2} \, \int_0^1 dt \, \Bigl[ \,
\langle \, V_3^{\xi \xi} (\xi \Psi, t \Psi, t \Psi) -V_3^{\xi \xi} (t \Psi, \xi \Psi, t \Psi)
+V_3^{\xi \xi} (t \Psi, t \Psi, \xi \Psi), Q t \Psi \, \rangle \\
& \qquad \qquad \qquad
+\langle \, V_2^\xi (V_2^\xi (\xi \Psi, t \Psi), t \Psi)-V_2^\xi (V_2^\xi (t \Psi, \xi \Psi), t \Psi)
+V_2^\xi (V_2^\xi (t \Psi, t \Psi), \xi \Psi), Q t \Psi \, \rangle \\
& \qquad \qquad \qquad
-\langle \, V_2^\xi (\xi \Psi, V_2^\xi (t \Psi, t \Psi))+V_2^\xi (t \Psi, V_2^\xi (\xi \Psi, t \Psi))
-V_2^\xi (t \Psi, V_2^\xi (t \Psi, \xi \Psi)), Q t \Psi \, \rangle \, \Bigr] \,.
\end{split}
\end{equation}
We find that the action is written in the form
\begin{equation}
S = {}-\frac{1}{g^2} \int_0^1 dt \, \langle \, A_t (t), Q A_\eta (t) \, \rangle
\end{equation}
with
\begin{equation}
\begin{split}
A_\eta (t)
& = g \, t \Psi +g^2 \, V_2^\xi (t \Psi, t \Psi) \\
& \quad~ +\frac{g^3}{2} \, \Bigl[ \, V_3^{\xi \xi} (t \Psi, t \Psi, t \Psi)
+V_2^\xi (V_2^\xi (t \Psi, t \Psi), t \Psi)
+V_2^\xi (t \Psi, V_2^\xi (t \Psi, t \Psi)) \, \Bigr] +O(g^4) \,, \\
A_t (t)
& = g \, \xi \Psi -g^2 \, \Bigl[ \, V_2^\xi (\xi \Psi, t \Psi) -V_2^\xi (t \Psi, \xi \Psi) \, \Bigr] \\
& \quad~ +\frac{g^3}{2} \, \Bigl[ \,
V_3^{\xi \xi} (\xi \Psi, t \Psi, t \Psi) -V_3^{\xi \xi} (t \Psi, \xi \Psi, t \Psi)
+V_3^{\xi \xi} (t \Psi, t \Psi, \xi \Psi) \\
& \qquad \qquad~
+V_2^\xi (V_2^\xi (\xi \Psi, t \Psi), t \Psi) -V_2^\xi (V_2^\xi (t \Psi, \xi \Psi), t \Psi)
+V_2^\xi (V_2^\xi (t \Psi, t \Psi), \xi \Psi) \\
& \qquad \qquad~
-V_2^\xi (\xi \Psi, V_2^\xi (t \Psi, t \Psi)) -V_2^\xi (t \Psi, V_2^\xi (\xi \Psi, t \Psi))
+V_2^\xi (t \Psi, V_2^\xi (t \Psi, \xi \Psi)) \, \Bigr] +O(g^4) \,.
\end{split}
\end{equation}

Let us next write the action in terms of $\Psi (t)$ following subsection~\ref{topological}.
For $\Psi (t)$ we define $A_\eta (t)$ and $A_t (t)$ by
\begin{align}
A_\eta (t)
& = g \, \Psi (t) +g^2 \, V_2^\xi ( \Psi (t), \Psi (t) ) \nonumber \\
& \quad~ +\frac{g^3}{2} \, \Bigl[ \, V_3^{\xi \xi} ( \Psi (t), \Psi (t), \Psi (t) )
+V_2^\xi ( V_2^\xi ( \Psi (t), \Psi (t) ), \Psi (t) )
+V_2^\xi ( \Psi (t), V_2^\xi ( \Psi (t), \Psi (t) ) ) \, \Bigr] \nonumber \\
& \quad~ +O(g^4)
\end{align}
and
\begin{align}
A_t (t)
& = g \, \xi \partial_t \Psi (t)
-g^2 \, \Bigl[ \, V_2^\xi ( \xi \partial_t \Psi (t), \Psi (t) )
-V_2^\xi ( \Psi (t), \xi \partial_t \Psi (t) ) \, \Bigr] \nonumber \\
& \quad~ +\frac{g^3}{2} \, \Bigl[ \,
V_3^{\xi \xi} ( \xi \partial_t \Psi (t), \Psi (t), \Psi (t) )
-V_3^{\xi \xi} ( \Psi (t), \xi \partial_t \Psi (t), \Psi (t) )
+V_3^{\xi \xi} ( \Psi (t), \Psi (t), \xi \partial_t \Psi (t) ) \nonumber \\
& \qquad \qquad~
+V_2^\xi (V_2^\xi ( \xi \partial_t \Psi (t), \Psi (t) ), \Psi (t) )
-V_2^\xi (V_2^\xi ( \Psi (t), \xi \partial_t \Psi (t) ), \Psi (t) ) \nonumber \\
& \qquad \qquad~
+V_2^\xi (V_2^\xi ( \Psi (t), \Psi (t) ), \xi \partial_t \Psi (t) )
-V_2^\xi ( \xi \partial_t \Psi (t), V_2^\xi ( \Psi (t), \Psi (t) ) ) \nonumber \\
& \qquad \qquad~
-V_2^\xi ( \Psi (t), V_2^\xi ( \xi \partial_t \Psi (t), \Psi (t) ) )
+V_2^\xi ( \Psi (t), V_2^\xi ( \Psi (t), \xi \partial_t \Psi (t) ) ) \, \Bigr] \nonumber \\
& \qquad \qquad~ +O(g^4) \,.
\end{align}
These coincide with
\begin{equation}
\begin{split}
A_\eta (t) & = \pi_1 \, {\bf G} \, \frac{1}{1-g \, \Psi (t)} \\
& = g \, \Psi (t)
+g^2 \, {\bm \mu}_2 \, ( \, \Psi (t) \otimes \Psi (t) \, ) \\
& \quad~
+\frac{g^3}{2} \, \Bigl[ \,
{\bm \mu}_3 \, ( \, \Psi (t) \otimes \Psi (t) \otimes \Psi (t) \, )
+{\bm \mu}_2 \, {\bm \mu}_2 \, ( \, \Psi (t) \otimes \Psi (t) \otimes \Psi (t) \, ) \, \Bigr]
+O(g^4)
\end{split}
\end{equation}
and
\begin{align}
& A_t (t) = \pi_1 {\bf G} \, {\bm \xi}_t \frac{1}{1-g \, \Psi (t)}
= \pi_1 {\bf G} \,
\Bigl( \, \frac{1}{1-g \, \Psi (t)} \otimes g \, \xi \partial_t \Psi (t) \otimes \frac{1}{1-g \, \Psi (t)} \, \Bigr)
\nonumber \\
& = g \, \xi \partial_t \Psi (t)
+g^2 \, {\bm \mu}_2 \, ( \, \xi \partial_t \Psi (t) \otimes \Psi (t)
+\Psi (t) \otimes \xi \partial_t \Psi (t) \, )
\nonumber \\
& \quad~
+\frac{g^3}{2} \, \Bigl[ \,
{\bm \mu}_3 \, ( \, \xi \partial_t \Psi (t) \otimes \Psi (t) \otimes \Psi (t)
+\Psi (t) \otimes \xi \partial_t \Psi (t) \otimes \Psi (t)
+\Psi (t) \otimes \Psi (t) \otimes \xi \partial_t \Psi (t) \, )
\nonumber \\
& \qquad \qquad
+{\bm \mu}_2 \, {\bm \mu}_2 \, ( \, \xi \partial_t \Psi (t) \otimes \Psi (t) \otimes \Psi (t)
+\Psi (t) \otimes \xi \partial_t \Psi (t) \otimes \Psi (t)
+\Psi (t) \otimes \Psi (t) \otimes \xi \partial_t \Psi (t) \, ) \, \Bigr]
\nonumber \\
& \quad~ +O(g^4)
\end{align}
under the translation in the preceding subsection.
From these expressions, we can explicitly confirm that
\begin{equation}
\begin{split}
\eta A_\eta (t) & = A_\eta (t) A_\eta (t) \,, \\
\partial_t A_\eta (t) & = \eta A_t (t) -A_\eta (t) A_t (t) +A_t (t) A_\eta (t)
\end{split}
\end{equation}
are satisfied up to $O(g^4)$.

\subsection{Field redefinition}

We define $\widetilde{A}_\eta (t)$ by
\begin{equation}
\widetilde{A}_\eta (t)
= ( \eta \, e^{g \, \xi \widetilde{\Psi} (t)} ) \, e^{-g \, \xi \widetilde{\Psi} (t)} \,.
\end{equation}
It can be expanded in $g$ as
\begin{equation}
\widetilde{A}_\eta (t)
= g \, \widetilde{\Psi} (t) +\frac{g^2}{2} \, [ \, \xi \widetilde{\Psi} (t), \widetilde{\Psi} (t) \, ]
+\frac{g^3}{6} \, [ \, \xi \widetilde{\Psi} (t), [ \, \xi \widetilde{\Psi} (t), \widetilde{\Psi} (t) \, ] \, ]
+O(g^4) \,.
\end{equation}
The string fields $\Psi (t)$ and $\widetilde{\Psi} (t)$ are related by
\begin{equation}
A_\eta (t) = \widetilde{A}_\eta (t) \,.
\end{equation}
We expand both sides in $g$ to obtain
\begin{align}
& g \, \Psi (t) +g^2 \, V_2^\xi ( \Psi (t), \Psi (t) ) \nonumber \\
& +\frac{g^3}{2} \, \Bigl[ \, V_3^{\xi \xi} ( \Psi (t), \Psi (t), \Psi (t) )
+V_2^\xi ( V_2^\xi ( \Psi (t), \Psi (t) ), \Psi (t) )
+V_2^\xi ( \Psi (t), V_2^\xi ( \Psi (t), \Psi (t) ) ) \, \Bigr] +O(g^4) \nonumber \\
& = g \, \widetilde{\Psi} (t) +\frac{g^2}{2} \, [ \, \xi \widetilde{\Psi} (t), \widetilde{\Psi} (t) \, ]
+\frac{g^3}{6} \, [ \, \xi \widetilde{\Psi} (t), [ \, \xi \widetilde{\Psi} (t), \widetilde{\Psi} (t) \, ] \, ]
+O(g^4) \,.
\end{align}
Let us expand the field redefinition from $\widetilde{\Psi} (t)$ to $\Psi (t)$ as
\begin{equation}
\Psi (t) = f^{(1)} +g \, f^{(2)} +g^2 f^{(3)} +O(g^3) \,.
\end{equation}
We then have
\begin{equation}
\begin{split}
& f^{(1)} = \widetilde{\Psi} (t) \,, \\
& f^{(2)} +V_2^\xi ( f^{(1)}, f^{(1)} \, )
= \frac{1}{2} \, [ \, \xi \widetilde{\Psi} (t), \widetilde{\Psi} (t) \, ] \,, \\
& f^{(3)} +V_2^\xi ( f^{(2)}, f^{(1)} \, )
+V_2^\xi ( f^{(1)}, f^{(2)} \, ) \\
& +\frac{1}{2} \, \Bigl[ \,
V_3^{\xi \xi} ( f^{(1)}, f^{(1)}, f^{(1)} \, )
+V_2^\xi ( V_2^\xi ( f^{(1)}, f^{(1)} ), f^{(1)} \, )
+V_2^\xi ( f^{(1)}, V_2^\xi ( f^{(1)}, f^{(1)} \, ) )
 \, \Bigr] \\
& = \frac{1}{6} \, [ \, \xi \widetilde{\Psi} (t), [ \, \xi \widetilde{\Psi} (t), \widetilde{\Psi} (t) \, ] \, ] \,.
\end{split}
\end{equation}
The field redefinition is given by
\begin{equation}
\begin{split}
f^{(1)} & = \widetilde{\Psi} (t) \,, \\
f^{(2)}
& = \frac{1}{2} \, [ \, \xi \widetilde{\Psi} (t), \widetilde{\Psi} (t) \, ]
-V_2^\xi ( \widetilde{\Psi} (t), \widetilde{\Psi} (t) ) \,, \\
f^{(3)}
& = \frac{1}{6} \, [ \, \xi \widetilde{\Psi} (t), [ \, \xi \widetilde{\Psi} (t), \widetilde{\Psi} (t) \, ] \, ]
-\frac{1}{2} \, V_2^\xi ( \, [ \, \xi \widetilde{\Psi} (t), \widetilde{\Psi} (t) \, ], \widetilde{\Psi} (t) )
-\frac{1}{2} \, V_2^\xi ( \widetilde{\Psi} (t), [ \, \xi \widetilde{\Psi} (t), \widetilde{\Psi} (t) \, ] \, ) \\
& \quad~ -\frac{1}{2} \, V_3^{\xi \xi} ( \widetilde{\Psi} (t), \widetilde{\Psi} (t), \widetilde{\Psi} (t) )
+\frac{1}{2} \, V_2^\xi ( V_2^\xi ( \widetilde{\Psi} (t), \widetilde{\Psi} (t) ), \widetilde{\Psi} (t) )
+\frac{1}{2} \, V_2^\xi ( \widetilde{\Psi} (t), V_2^\xi ( \widetilde{\Psi} (t), \widetilde{\Psi} (t) ) ) \,.
\end{split}
\end{equation}

The string field $A_t (t)$ written in terms of $\widetilde{\Psi} (t)$ is
\begin{equation}
A_t (t) = g \, A_t^{(1)} (t) +g^2 \, A_t^{(2)} (t) +g^3 \, A_t^{(3)} (t) +O(g^4) \,,
\end{equation}
where
\begin{equation}
\begin{split}
A_t^{(1)} (t) & = \xi \partial_t \widetilde{\Psi} (t) \,, \\
A_t^{(2)} (t)
& = \frac{1}{2} \, \xi \partial_t \, [ \, \xi \widetilde{\Psi} (t), \widetilde{\Psi} (t) \, ]
-\xi \partial_t V_2^\xi ( \widetilde{\Psi} (t), \widetilde{\Psi} (t) )
-V_2^\xi ( \xi \partial_t \Psi (t), \Psi (t) )
+V_2^\xi ( \Psi (t), \xi \partial_t \Psi (t) ) \,,
\end{split}
\end{equation}
and
\begin{align}
A_t^{(3)} (t)
& = \frac{1}{6} \, \xi \partial_t \, [ \, \xi \widetilde{\Psi} (t), [ \, \xi \widetilde{\Psi} (t), \widetilde{\Psi} (t) \, ] \, ]
-\frac{1}{2} \, \xi \partial_t \, V_2^\xi ( \, [ \, \xi \widetilde{\Psi} (t), \widetilde{\Psi} (t) \, ], \widetilde{\Psi} (t) )
\nonumber \\
& \quad~
-\frac{1}{2} \, \xi \partial_t \, V_2^\xi ( \widetilde{\Psi} (t), [ \, \xi \widetilde{\Psi} (t), \widetilde{\Psi} (t) \, ] \, )
-\frac{1}{2} \, \xi \partial_t \, V_3^{\xi \xi} ( \widetilde{\Psi} (t), \widetilde{\Psi} (t), \widetilde{\Psi} (t) )
\nonumber \\
& \quad~
+\frac{1}{2} \, \xi \partial_t \, V_2^\xi ( V_2^\xi ( \widetilde{\Psi} (t), \widetilde{\Psi} (t) ), \widetilde{\Psi} (t) )
+\frac{1}{2} \, \xi \partial_t \, V_2^\xi ( \widetilde{\Psi} (t), V_2^\xi ( \widetilde{\Psi} (t), \widetilde{\Psi} (t) ) )
\nonumber \\
& \quad~
{}-\frac{1}{2} \, V_2^\xi ( \xi \partial_t \, [ \, \xi \widetilde{\Psi} (t), \widetilde{\Psi} (t) \, ], \widetilde{\Psi} (t) )
+V_2^\xi ( \xi \partial_t V_2^\xi ( \widetilde{\Psi} (t), \widetilde{\Psi} (t) ), \widetilde{\Psi} (t) )
\nonumber \\
& \quad~
{}-\frac{1}{2} \, V_2^\xi ( \xi \partial_t \widetilde{\Psi} (t), [ \, \xi \widetilde{\Psi} (t), \widetilde{\Psi} (t) \, ] )
+V_2^\xi ( \xi \partial_t \widetilde{\Psi} (t), V_2^\xi ( \widetilde{\Psi} (t), \widetilde{\Psi} (t) ) )
\nonumber \\
& \quad~
+\frac{1}{2} \, V_2^\xi ( [ \, \xi \widetilde{\Psi} (t), \widetilde{\Psi} (t) \, ], \xi \partial_t \widetilde{\Psi} (t) )
-V_2^\xi ( V_2^\xi ( \widetilde{\Psi} (t), \widetilde{\Psi} (t) ), \xi \partial_t \widetilde{\Psi} (t) )
\nonumber \\
& \quad~
+\frac{1}{2} \, V_2^\xi ( \widetilde{\Psi} (t), \xi \partial_t \, [ \, \xi \widetilde{\Psi} (t), \widetilde{\Psi} (t) \, ] )
-V_2^\xi ( \widetilde{\Psi} (t), \xi \partial_t V_2^\xi ( \widetilde{\Psi} (t), \widetilde{\Psi} (t) ) )
\nonumber \\
& \quad~
+\frac{1}{2} \, V_3^{\xi \xi} ( \xi \partial_t \widetilde{\Psi} (t), \widetilde{\Psi} (t), \widetilde{\Psi} (t) )
-\frac{1}{2} \, V_3^{\xi \xi} ( \widetilde{\Psi} (t), \xi \partial_t \widetilde{\Psi} (t), \widetilde{\Psi} (t) )
\nonumber \\
& \quad~
+\frac{1}{2} \, V_3^{\xi \xi} ( \widetilde{\Psi} (t), \widetilde{\Psi} (t), \xi \partial_t \widetilde{\Psi} (t) )
\nonumber \\
& \quad~
+\frac{1}{2} \, V_2^\xi (V_2^\xi ( \xi \partial_t \widetilde{\Psi} (t), \widetilde{\Psi} (t) ), \widetilde{\Psi} (t) )
-\frac{1}{2} \, V_2^\xi (V_2^\xi ( \widetilde{\Psi} (t), \xi \partial_t \widetilde{\Psi} (t) ), \widetilde{\Psi} (t) )
\nonumber \\
& \quad~
+\frac{1}{2} \, V_2^\xi (V_2^\xi ( \widetilde{\Psi} (t), \widetilde{\Psi} (t) ), \xi \partial_t \widetilde{\Psi} (t) )
-\frac{1}{2} \, V_2^\xi ( \xi \partial_t \widetilde{\Psi} (t), V_2^\xi ( \widetilde{\Psi} (t), \widetilde{\Psi} (t) ) )\nonumber \\
& \quad~
-\frac{1}{2} \, V_2^\xi ( \widetilde{\Psi} (t), V_2^\xi ( \xi \partial_t \widetilde{\Psi} (t), \widetilde{\Psi} (t) ) )
+\frac{1}{2} \, V_2^\xi ( \widetilde{\Psi} (t), V_2^\xi ( \widetilde{\Psi} (t), \xi \partial_t \widetilde{\Psi} (t) ) ) \,. \end{align}
On the other hand, $\widetilde{A}_t (t)$ is expanded as
\begin{equation}
\widetilde{A}_t (t) = ( \partial_t \, e^{g \, \xi \widetilde{\Psi} (t)} ) \, e^{-g \, \xi \widetilde{\Psi} (t)}
= g \, \widetilde{A}_t^{\, (1)} (t) +g^2 \, \widetilde{A}_t^{\, (2)} (t)
+g^3 \, \widetilde{A}_t^{\, (3)} (t) +O(g^4) \,,
\end{equation}
where
\begin{equation}
\begin{split}
\widetilde{A}_t^{\, (1)} (t) & = \xi \partial_t \widetilde{\Psi} (t) \,, \\
\widetilde{A}_t^{\, (2)} (t) 
& = \frac{1}{2} \, [ \, \xi \widetilde{\Psi} (t), \xi \partial_t \widetilde{\Psi} (t) \, ] \,, \\
\widetilde{A}_t^{\, (3)} (t) 
& = \frac{1}{6} \,
[ \, \xi \widetilde{\Psi} (t), [ \, \xi \widetilde{\Psi} (t), \xi \partial_t \widetilde{\Psi} (t) \, ] \, ] \,.
\end{split}
\end{equation}
We find
\begin{equation}
A_t (t) \ne \widetilde{A}_t (t) \,,
\end{equation}
and the difference is
\begin{equation}
\begin{split}
\Delta A_t (t) & = A_t (t) -\widetilde{A}_t (t)
= g^2 \, ( \, A_t^{(2)} (t) -\widetilde{A}_t^{\, (2)} (t) \, )
+g^3 \, ( \, A_t^{(3)} (t) -\widetilde{A}_t^{\, (3)} (t) \, ) +O(g^4) \,.
\end{split}
\end{equation}
In subsection~\ref{field-redefinition}, we have shown
that the difference $\Delta A_t (t)$ is annihilated
by the covariant derivative $D_\eta (t)$.
Let us confirm this explicitly to $O(g^4)$. Since
\begin{equation}
\begin{split}
D_\eta (t) \, \Delta A_t (t) & = g^2 \, \eta \, ( A_t^{(2)} (t) -\widetilde{A}_t^{\, (2)} (t) ) \\
& \quad~
+g^3 \, \Bigl( \, \eta \, ( A_t^{(3)} (t) -\widetilde{A}_t^{\, (3)} (t) )
-[ \, \widetilde{\Psi} (t) \,,\, A_t^{(2)} (t) -\widetilde{A}_t^{\, (2)} (t) \, ] \, \Bigr) +O(g^4) \,,
\end{split}
\end{equation}
we need to show
\begin{align}
\eta \, ( A_t^{(2)} (t) -\widetilde{A}_t^{\, (2)} (t) ) & = 0 \,,
\label{D_eta-A_t-g^2} \\
\eta \, ( A_t^{(3)} (t) -\widetilde{A}_t^{\, (3)} (t) )
-[ \, \widetilde{\Psi} (t) \,,\, A_t^{(2)} (t) -\widetilde{A}_t^{\, (2)} (t) \, ] & = 0 \,.
\label{D_eta-A_t-g^3}
\end{align}
For $\eta A_t^{(2)} (t)$ and $\eta \widetilde{A}_t^{\, (2)} (t)$, we find
\begin{equation}
\begin{split}
\eta A_t^{(2)} (t)
& = \frac{1}{2} \, \partial_t \, [ \, \xi \widetilde{\Psi} (t), \widetilde{\Psi} (t) \, ]
-\frac{1}{2} \, \xi \partial_t \, \{ \, \widetilde{\Psi} (t), \widetilde{\Psi} (t) \, \}
-\partial_t V_2^\xi ( \widetilde{\Psi} (t), \widetilde{\Psi} (t) )
+\xi \partial_t \, ( \widetilde{\Psi} (t) \, \widetilde{\Psi} (t) ) \\
& \quad~ -( \xi \partial_t \Psi (t) ) \, \Psi (t)
+V_2^\xi ( \partial_t \Psi (t), \Psi (t) )
+\Psi (t) \, ( \xi \partial_t \Psi (t) )
+V_2^\xi ( \Psi (t), \partial_t \Psi (t) ) \\
& = \frac{1}{2} \, \partial_t \, [ \, \xi \widetilde{\Psi} (t), \widetilde{\Psi} (t) \, ]
+[ \, \Psi (t), \xi \partial_t \Psi (t) \, ]
= \frac{1}{2} \, [ \, \xi \widetilde{\Psi} (t), \partial_t \widetilde{\Psi} (t) \, ]
+\frac{1}{2} \, [ \, \Psi (t), \xi \partial_t \Psi (t) \, ] \,,
\end{split}
\end{equation}
and
\begin{equation}
\begin{split}
\eta \widetilde{A}_t^{\, (2)} (t)
= \frac{1}{2} \, [ \, \widetilde{\Psi} (t), \xi \partial_t \widetilde{\Psi} (t) \, ]
+\frac{1}{2} \, [ \, \xi \widetilde{\Psi} (t), \partial_t \widetilde{\Psi} (t) \, ] \,.
\end{split}
\end{equation}
We have thus confirmed~\eqref{D_eta-A_t-g^2}.
The calculation of $\eta A_t^{(3)} (t)$ is tedious but straightforward. The upshot is
\begin{equation}
\eta A_t^{(3)} (t)
= \frac{1}{6} \, \partial_t \, [ \, \xi \widetilde{\Psi} (t), [ \, \xi \widetilde{\Psi} (t), \widetilde{\Psi} (t) \, ] \, ]
-\frac{1}{2} \, [ \, \xi \partial_t \widetilde{\Psi} (t), \, [ \, \xi \widetilde{\Psi} (t), \widetilde{\Psi} (t) \, ] \, ]
+[ \, \widetilde{\Psi} (t), A_t^{(2)} (t) \, ] \,.
\end{equation}
For $\eta \widetilde{A}_t^{\, (3)} (t)$, we have
\begin{equation}
\eta \widetilde{A}_t^{\, (3)} (t) 
= \frac{1}{6} \,
[ \, \widetilde{\Psi} (t), [ \, \xi \widetilde{\Psi} (t), \xi \partial_t \widetilde{\Psi} (t) \, ] \, ]
+\frac{1}{6} \,
[ \, \xi \widetilde{\Psi} (t), [ \, \widetilde{\Psi} (t), \xi \partial_t \widetilde{\Psi} (t) \, ] \, ]
+\frac{1}{6} \,
[ \, \xi \widetilde{\Psi} (t), [ \, \xi \widetilde{\Psi} (t), \partial_t \widetilde{\Psi} (t) \, ] \, ] \,.
\end{equation}
It is not difficult to verify that the relation~\eqref{D_eta-A_t-g^3} is satisfied.

Finally, let us write $Q A_\eta (t)$ in the form
\begin{equation}
Q A_\eta (t) = {}-D_\eta (t) \, A_Q (t)
\label{Q-A_eta-appendix}
\end{equation}
up to $O(g^3)$. Since
\begin{equation}
\begin{split}
& \eta \, \Bigl( \,
g \, \xi Q \Psi +g^2 \, \xi V_2 ( \Psi, \Psi )
+g^2 \, V_2^\xi ( \xi Q \Psi, \Psi ) +g^2 \, V_2^\xi ( \Psi, \xi Q \Psi ) \, \Bigr) \\
& = g \, Q \Psi +g^2 \, V_2 ( \Psi, \Psi )
+g^2 \, ( \xi Q \Psi ) \, \Psi -g^2 \, V_2^\xi ( Q \Psi, \Psi ) 
+g^2 \, \Psi \, ( \xi Q \Psi ) +g^2 \, V_2^\xi ( \Psi, Q \Psi ) \\
& = g \, Q \Psi +g^2 \, Q V_2^\xi ( \Psi, \Psi ) +g^2 \, \{ \, \Psi, \xi Q \Psi \, \} \,,
\end{split}
\end{equation}
we have
\begin{equation}
\begin{split}
& Q A_\eta (t) = g \, Q \Psi +g^2 \, Q V_2^\xi ( \Psi, \Psi ) +O(g^3) \\
& = \eta \, \Bigl( \,
g \, \xi Q \Psi +g^2 \, \xi V_2 ( \Psi, \Psi )
+g^2 \, V_2^\xi ( \xi Q \Psi, \Psi ) +g^2 \, V_2^\xi ( \Psi, \xi Q \Psi ) \, \Bigr)
-g^2 \, \{ \, \Psi, \xi Q \Psi \, \} +O(g^3) \,.
\end{split}
\end{equation}
This takes the form~\eqref{Q-A_eta-appendix} with
\begin{equation}
A_Q (t) = {}-g \, \xi Q \Psi -g^2 \, \Bigl( \, \xi V_2 ( \Psi, \Psi )
+V_2^\xi ( \xi Q \Psi, \Psi ) +V_2^\xi ( \Psi, \xi Q \Psi ) \, \Bigr) +O(g^3) \,.
\end{equation}

\end{document}